\documentclass[aps,pra,twocolumn,superscriptaddress,notitlepage]{revtex4-2}

\usepackage{soul}
\usepackage{relsize}
\usepackage{slantsc}
\usepackage{graphicx}
\usepackage{amsmath}
\usepackage{amssymb}
\usepackage{mathtools}
\usepackage{amsthm}
\usepackage{amsbsy}
\usepackage{mathrsfs}
\usepackage{varioref}
\usepackage{dsfont}
\usepackage{bm}
\usepackage{color}
\usepackage[usenames,dvipsnames]{xcolor}
\definecolor{myblue}{rgb}{0.2,0.2,0.8}
\definecolor{myblack}{rgb}{0,0,0}
\definecolor{myurl}{rgb}{0.1,0.1,0.4}
\usepackage[colorlinks=true,citecolor=myblue,linkcolor=myblack,urlcolor=myurl]{hyperref}
\usepackage{cleveref}
\usepackage[font = footnotesize, labelfont = bf, justification = RaggedRight]{caption}
\usepackage{subfigure}
\usepackage{booktabs} 
\usepackage{array} 
\usepackage{paralist} 
\usepackage{verbatim} 
\edef\restoreparindent{\parindent=\the\parindent\relax}
\usepackage[parfill]{parskip}
\usepackage{natbib}

\newcommand{\<}{\langle}
\renewcommand{\>}{\rangle}

\newcommand{\ww}{{\mathcal{W}}}
\newcommand{\kk}{{\mathcal{K}}}

\newcommand{\lo}{{\mathcal{L}}}

\newcommand{\trc}{{\mathcal{T}}}
\newcommand{\s}{{\mathcal{S}}}
\newcommand{\e}{{\mathcal{E}}}

\newcommand{\xx}{{\mathcal{X}}}

\renewcommand{\aa}{{\mathcal{A}}}

\newcommand{\dd}{{\mathcal{D}}}

\newcommand{\qq}{{\mathcal{Q}}}
\newcommand{\mm}{{\mathcal{M}}}

\newcommand{\ii}{{\mathcal{I}}}
\newcommand{\jj}{{\mathcal{J}}}

\newcommand{\h}{{\mathcal{H}}}

\newcommand{\ha}{{\mathcal{H}\sub{\aa}}}
\newcommand{\hs}{{\mathcal{H}\sub{\s}}}

\newcommand{\E}{\mathsf{E}}
\newcommand{\Z}{\mathsf{Z}}

\newcommand{\co}{\mathds{C}}
\newcommand{\re}{\mathds{R}}

\newcommand{\one}{\mathds{1}}
\newcommand{\onesys}{\mathds{1}\sub{\s}}
\newcommand{\oneapp}{\mathds{1}\sub{\aa}}
\newcommand{\zero}{\mathds{O}}

\newcommand{\idchsys}{\mathbb{I}\sub{\s}}

\newcommand{\qv}{\mathrm{V}\sub{\mathrm{qu}}}
\newcommand{\cv}{\mathrm{V}\sub{\mathrm{cl}}}

\newcommand{\tr}{\mathrm{tr}}

\newcommand{\imag}{\mathfrak{i}}

\newcommand{\sub}[1]{_{\!\mathsmaller{\, #1}}}
\newcommand{\eq}[1]{Eq.~\eqref{#1}}
\newcommand{\fig}[1]{Fig.~\ref{#1}}
\newcommand{\sect}[1]{Sec.~\ref{#1}}
\newcommand{\app}[1]{Appendix~(\ref{#1})}
\newcommand{\ket}[1]{|{#1}\rangle}

\newcommand{\avg}[1]{\langle {#1} \rangle} 
\newcommand{\var}[1]{\mathrm{Var}\left({#1} \right)}

\begin{document}

\title{Classicality of the heat produced by quantum measurements}

\author{M. Hamed Mohammady}
\affiliation{QuIC, \'{E}cole Polytechnique de Bruxelles, CP 165/59, Universit\'{e} Libre de Bruxelles, 1050 Brussels, Belgium}
\affiliation{RCQI, Institute of Physics, Slovak Academy of Sciences, D\'ubravsk\'a cesta 9, Bratislava 84511, Slovakia}


\begin{abstract}
Quantum measurement is ultimately a physical process, resulting from an interaction between the measured system and a measuring apparatus. Considering the physical process of measurement within a thermodynamic context naturally raises the following question: How can the work and heat  be interpreted? In the present paper, we model the measurement process for an arbitrary discrete observable as a \emph{measurement scheme}. Here, the system to be measured is first unitarily coupled with an apparatus, and subsequently the compound system is \emph{objectified} with respect to a \emph{pointer observable}, thus producing  definite measurement outcomes. The work can therefore be interpreted as the change in internal energy of the compound system due to the unitary coupling. By the first law of thermodynamics, the heat is the subsequent change in internal energy of this compound due to pointer objectification. We argue that the apparatus serves as a stable record for the measurement outcomes only if  the  pointer observable commutes with the Hamiltonian,  and  show that such commutativity implies that the uncertainty of heat will necessarily be  classical. 
\end{abstract}

\maketitle

\section{Introduction} 
Quantum measurements play a central role in quantum thermodynamics: they are used in several formulations of fluctuation relations \cite{Funo2013, Roncaglia2014, Watanabe2014, Perarnau-Llobet2016a, Morikuni2017, Elouard2017b, DeChiara2018, Sone2020, Mohammady2020,  Hovhannisyan2021}, and  they fuel quantum thermal machines \cite{Jacobs2009, Shiraishi2015a, Hayashi2017,Mohammady2017, Elouard2017a, Elouard2018b, Manzano2018, Buffoni2018, Solfanelli2019,Purves-2020, Bresque2020}.  The thermodynamic properties of the measurement process has  also been a subject of  investigation \cite{Zurek2003a,  Miyadera2011d,  Navascues2014a, Miyadera2015a, Allahverdyan2013a,  Konishi2016a,   Mancino2017,Benoist2017, Guryanova2018, Benoist2020,Naikoo2021,Landi-steady-state},   with the interpretation of the work and heat that result from measurement being a hotly debated topic. The quantity that is most commonly considered in this regard is the  energy that is dissipated to a thermal environment as a result of erasing the record of measurement outcomes stored in the measurement apparatus \cite{Sagawa2009b, Jacobs2012a, Abdelkhalek2016}. The lower bound to this quantity is the  Landauer erasure cost \cite{Landauer1961, Reeb2013a}---determined only by the entropy change of the apparatus and the temperature of the surrounding thermal environment---which can also be interpreted as the contribution from the  apparatus to the total non-recoverable work of the measurement process \cite{Mohammady2019c}.

However, erasure occurs after the measurement process is completed, and therefore strictly speaking is independent of the measurement process as such.  An interpretation of the work and heat  resulting from the measurement process itself,  and    independent of the erasure process that takes place after measurement, was  recently given by Strasberg \cite{Strasberg2019}. Here, an observable is ``indirectly'' measured  by means of a \emph{measurement scheme} \cite{Ozawa1984}, where  the system to be measured is first unitarily coupled with a quantum probe, and thereafter the probe is measured by a \emph{pointer observable}. The registered  outcomes of the pointer observable are in a one-to-one relation with the outcomes of the system observable, and are detected by the same probability as if the system observable were measured ``directly''.   The work for such a measurement process was  identified as the change in internal energy of the compound of system-plus-probe due to their unitary evolution. By the first law of thermodynamics, the heat was thus shown to be the subsequent change in internal energy of the compound when the measurement of the pointer observable registers a given outcome. Such heat results from the state change that accompanies quantum measurements, and is thus an intrinsically stochastic quantity; indeed, this  definition for  heat is similar to the so-called ``quantum heat'' introduced by  Elouard et al  \cite{Elouard2017},  defined as  the change in internal energy of  only the measured system given that a  measurement outcome has been observed.  

An indirect measurement scheme implicitly assumes that the external observer has access to a macroscopic measurement apparatus used to measure the probe by the pointer observable. Since the apparatus registers the definite measurement outcomes, the probe may be discarded after the measurement process has been completed. Therefore,  the  state change of the probe caused by the measurement of the pointer observable is unimportant insofar as the measurement statistics of the system observable is concerned.  In the present manuscript, however,  we shall  consider   measurement schemes as a quantum mechanical model for a  ``direct'' measurement, where the probe is treated as a quantum mechanical representation of the apparatus itself. As with indirect measurement schemes, we identify work with the initial unitary interaction between system and apparatus---a process referred to as \emph{premeasurement}. On the other hand, heat is identified with the subsequent pointer \emph{objectification}, that is, the process by which the   compound of system-plus-apparatus is transformed into a state for which the pointer observable takes definite (objective) values, which can then be ``read'' by the observer without causing further disturbance \cite{Busch-Shimony-1997}.  The possibility of objectification therefore demands that all effects of the pointer observable must have at least one eigenvector with eigenvalue 1, with the objectified states of the compound system having support only in these eigenvalue-1 eigenspaces. While a sharp pointer observable trivially satisfies this requirement, we will consider the  more general case where the pointer observable can be unsharp \cite{Busch1998a}.    However, in order for the apparatus to serve  as a stable record  of the measurement outcomes,   we demand that the pointer observable must commute with the Hamiltonian;  we refer to the commutation between the pointer observable and the Hamiltonian as the \emph{Yanase condition} \cite{Yanase1961, Ozawa2002}, which was first introduced in the context of the  Wigner-Araki-Yanase theorem \cite{E.Wigner1952, Busch2010, Araki1960}. 

The Yanase condition follows from the fact that if work is to be fully identified with the initial unitary interaction between system and apparatus, the  compound system must be governed by a time-independent Hamiltonian during (and after) pointer objectification. Such time-independence of the Hamiltonian implies that the only pointer observables that can be measured are those that are invariant under time-translation symmetry, i.e., observables that commute with the Hamiltonian \cite{Loveridge2017a, Loveridge2020a}. Indeed, if the pointer observable does not commute with the Hamiltonian, then the outcome revealed by the observer's  measurement of the pointer observable after objectification will be time-dependent; the record of the measurement outcome will not be stable. We show that for measurement schemes where the pointer observable satisfies the Yanase condition, the uncertainty of the heat that results from objectification will necessarily be  classical. This is because the objectified states of system-plus-apparatus will be pairwise orthogonal and, together with the Yanase condition, such orthogonality guarantees that the quantum contribution to the  heat uncertainty---which is a function of the Wigner-Yanase-Dyson skew information \cite{Wigner1963, Lieb1973}---entirely vanishes.  The classicality of the heat uncertainty may be interpreted within information theoretic terms as reflecting the fact that the information content---both of the measurement outcomes and the time elapsed from objectification---stored in the objectified states of system-plus-apparatus is perfectly transmitted to the observer.

The manuscript is organised as follows: In \sect{sec:preliminaries} we review the basic elements of the quantum theory of measurement. In \sect{sec:measurement-scheme} we characterise measurement schemes as models for direct measurement processes, and in \sect{sec:work-heat} we evaluate the work and heat that results from the measurement process. In \sect{sec:Yanase} we argue for the necessity of the Yanase condition, and in \sect{sec:heat-uncertainty} show that the Yanase condition ensures classicality of the heat uncertainty. Finally, in \sect{sec:Luders} we give a concrete example where the system is measured by the L\"uders instrument.

\section{Quantum measurement}\label{sec:preliminaries}
In this section, we shall give a brief but self-contained review of  the quantum theory of measurement. For further details, we refer to the texts \cite{PaulBuschMarianGrabowski1995, Busch1996, PeterMittelstaedt2004, Heinosaari2011, Holevo-Prob-Quantum, Busch2016a,Hayashi-QIT}. 

\subsection{Basic concepts}
We consider systems with a separable complex Hilbert space $\h$, and shall denote  with $\lo(\h)$ the algebra of bounded linear operators on $\h$. $\zero$ and $\one$ will represent the      null and identity operators of $\lo(\h)$, respectively.   We further define by $\trc(\h) \subseteq \lo(\h)$ the space of trace-class operators, and by  $\s(\h) \subset \trc(\h)$  the  space  of positive unit-trace operators, i.e., states.   

In the Schr\"odinger picture, physical transformations  will be represented by \emph{operations}, that is, completely positive trace non-increasing linear maps $\Phi : \trc(\h) \to \trc(\kk)$, where $\h$ is the input space and $\kk$ the output.    Transformations in the Heisenberg picture  will be described by the  dual operations $\Phi^* : \lo(\kk) \to \lo(\h)$, which are completely positive sub-unital linear maps,  defined by the trace duality $\tr[\Phi^*(B) T] = \tr[ B \Phi(T)]$ for all $B \in \lo(\kk)$ and $T \in \trc(\h)$. The trace preserving (or unital)  operations  will be referred to as channels.

\subsection{Discrete observables}

At the coarsest level of description, an observable of a system with Hilbert space $\h$  is represented by, and identified with,  a normalised positive operator valued measure (POVM) $\E: \Sigma \to \lo(\h)$. $\Sigma$ is the $\sigma$-algebra of some value space $\xx$, and represents the possible measurement outcomes of $\E$. For any $X \in \Sigma$, the positive operator $\zero \leqslant \E(X) \leqslant \one$ is referred to as the associated  \emph{effect} of $\E$, and normalisation implies that $\E(\xx) = \one$. If the value space is a countable set   $\xx = \{x_1, x_2, \dots\}$, then $\E$ is referred to as a \emph{discrete} observable. Unless stated otherwise, we shall always assume that $\E$ is discrete, in which case   it can be identified by the set of effects as $\E := \{\E_x  \equiv \E(\{x\}) : x \in \xx \}$,  such that $\sum_{x\in \xx} \E_x  = \one$ (converging weakly). The probability of observing outcome $x$ when the  observable $\E$ is measured in the  state $\rho$ is given by the Born rule as
\begin{align*}
p^\E_\rho (x) := \tr[\E_x  \rho].
\end{align*}
An observable is called \emph{sharp} if all the effects are projection operators, i.e., if $\E_x  \E_y   = \delta_{x,y}\E_x $. Sharp observables correspond with self-adjoint operators by the spectral theorem. An observable that is not sharp will be referred to as \emph{unsharp}. An observable admits \emph{objective} values if for all $x\in \xx$, there exists a state $\rho$ which is \emph{objectified} with respect to $\E$, i.e.,  $\tr[\E_x \rho]=1$. This implies that all effects $\E_x$ must have at least one eigenvector with eigenvalue 1, where $\rho$ is objectified with respect to $\E$ if $\rho$ only has support in the eigenvalue-1 eigenspace of $\E_x$. Sharp observables trivially satisfy this condition, but so do certain unsharp observables.

\begin{figure}[htbp!]
\begin{center}
\includegraphics[width=0.47\textwidth]{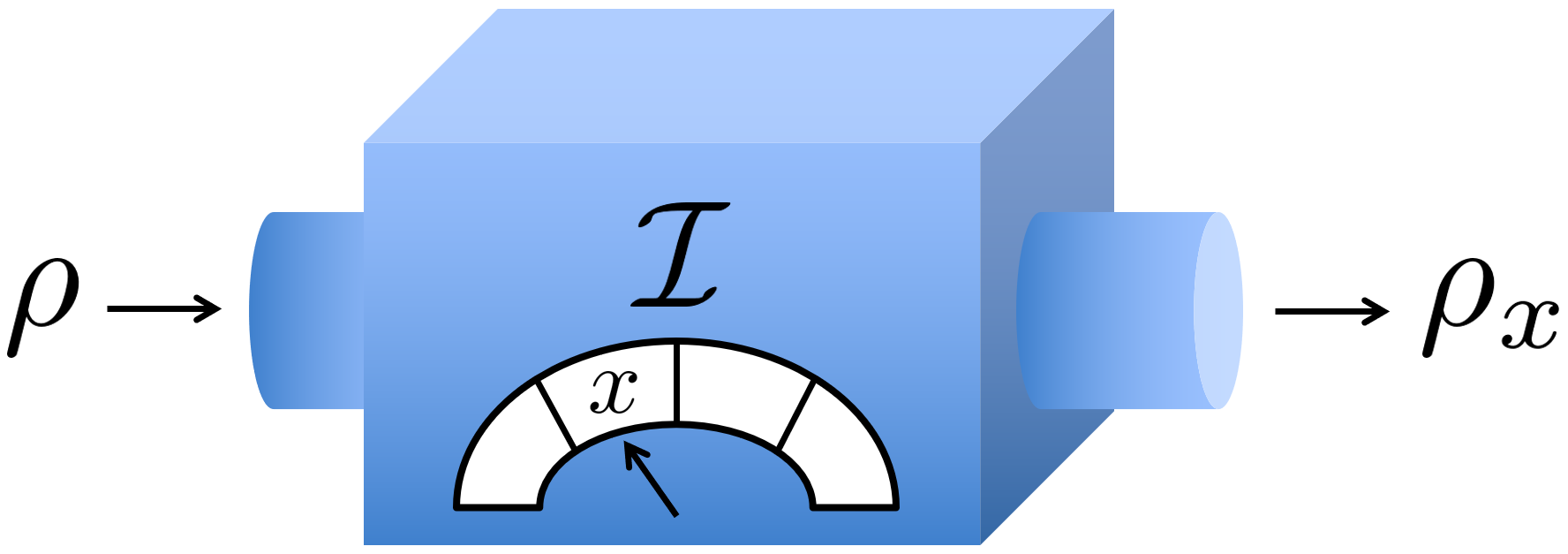}
\vspace*{-0.2cm}
\caption{A discrete observable $\E$ is measured by an instrument $\ii$. A quantum system, prepared in an arbitrary state $\rho$, enters the instrument which registers outcome $x$ with probability $p^\E_\rho (x)   := \tr[\E_x  \rho] = \tr[\ii_x(\rho)]$,    and transforms the system to the new state $\rho_x := \ii_x(\rho)/p^\E_\rho (x)$. }\label{fig:Fig1}
\vspace*{-0.5cm}
\end{center}
\end{figure}

\subsection{Instruments}

A more detailed representation of observables is given by instruments, or normalised operation valued  measures, which describe how  a measured system is transformed \cite{Davies1970}. A discrete instrument $\ii$ is fully characterised by the set of operations $\ii := \{\ii_x \equiv \ii(\{x\}) :  x\in \xx\}$, where  $\ii_x : \trc(\h) \to \trc(\h)$ are the operations of $\ii$, such that $\sum_{x\in \xx}\tr[\ii_x(T)] = \tr[T]$ for all $T \in \trc(\h)$. We shall therefore identify $\ii_\xx(\cdot) := \sum_{x\in \xx} \ii_x(\cdot)$ as the channel induced by $\ii$. An instrument $\ii$ is identified with a unique observable $\E$ via the relation  $ \ii_x^*(\one) = \E_x $ for all $x\in \xx$. This  implies that for all states and outcomes,  $p^\E_\rho (x) := \tr[\E_x  \rho] = \tr[\ii_x(\rho)] $. We shall refer to $\ii$ as being  \emph{compatible} with observable $\E$, or an $\E$-instrument for short.      Note that while an instrument $\ii$ is compatible with a unique observable $\E$, an observable  admits many different instruments. 

For an input state $\rho \in \s(\h)$, and any outcome $x$ such that $p^\E_\rho (x) >0$,  the normalised conditional state prepared by $\ii$ is defined as 
\begin{align}\label{eq:conditional-state}
\rho_x := \ii_x(\rho)/p^\E_\rho (x).
\end{align}
If $p^\E_\rho (x) =0$, we define $\rho_x := \zero$.  On the other hand, the unconditional state given a non-selective measurement is  $\ii_\xx(\rho)$.  \fig{fig:Fig1} gives a schematic representation of an instrument. 

 An $\E$-instrument $\ii$ is  \emph{repeatable}  if for all $x,y\in \xx$ and $\rho \in \s(\h)$,  
\begin{align}\label{eq:repeatable-defn-states}
\tr[\E_y   \ii_x(\rho)] = \delta_{x,y} \tr[\E_x  \rho],
\end{align}
which may equivalently be written as 
\begin{align}\label{eq:repeatable-defn}
 \ii_x^*(\E_y  ) = \delta_{x,y} \E_x  \, \forall \, x,y \in \xx.
\end{align}
In other words, $\ii$ is a repeatable $\E$-instrument   if a second measurement of $\E$ is guaranteed to produce the same outcome as  $\ii$. An observable $\E$ admits a  repeatable instrument only if it is discrete \cite{Ozawa1984}, and all of its effects have at least one eigenvector with eigenvalue 1  \cite{ Busch1995}. 

We note that the repeatability condition \eq{eq:repeatable-defn-states} is equivalent to $\tr[\E_x \ii_x(\rho)] = \tr[\E_x \rho]$ for all $x$ and $\rho$ \cite{Busch1990}. Therefore, by \eq{eq:conditional-state},     $\ii$ is a repeatable $\E$-instrument if and only if for all $x \in \xx$ and $\rho \in \s(\h)$ such that $p^\E_\rho (x)>0$,   
\begin{align*}
\tr[\E_x  \rho_x] =  \frac{\tr[\E_x   \ii_x(\rho) ]}{\tr[\E_x  \rho]} =  \frac{\tr[\E_x \rho ]}{\tr[\E_x  \rho]} = 1.
\end{align*}
 Let us denote the projection onto the eigenvalue-1 eigenspace of the effect $\E_x$ as $P_x$, where we note that since $\sum_x \E_x =\one$, then $P_x P_y = P_x \E_y = \delta_{x,y} P_x$.    Since  $\tr[\E_x \rho_x] = 1$ if and only if   $\tr[P_x \rho_x]=1$, then we may infer that the instrument $\ii$ is repeatable if and only if  for all input states $\rho$, the output states may be written as $\rho_x = P_x \rho_x P_x = P_x\rho_x =\rho_x P_x$; the output state $\rho_x$ must only have support in the eigenvalue-1 eigenspace of $\E_x$.  It is easily seen that if $\ii$ is repeatable, then for all input states $\rho$,  the output states  will be pairwise orthogonal, since  $\rho_x \rho_y = \rho_x P_x P_y \rho_y = \zero$ if $x\ne y$.  

All discrete observables $\E$ can be implemented by a L\"uders instrument $\ii^L$ defined as
\begin{align*}
\ii^{L}_x(T) := \sqrt{\E_x} \,  T \, \sqrt{\E_x},
\end{align*}
to hold for all $x\in \xx$ and $T\in \trc(\h)$ \cite{Luders2006}. A L\"uders instrument is repeatable if and only if $\E$ is sharp; noting that ${\ii^L_x}^* = \ii^L_x$, we see that  ${\ii^L_x}^*(\E_x) = \E_x^2$, which satisfies  \eq{eq:repeatable-defn} if and only if $\E_x^2 = \E_x$. But for any observable $\E$, all effects of which have at least one eigenvector with eigenvalue 1,  the L\"uders instrument implements an \emph{ideal} measurement of $\E$. Ideal measurements only disturb the system to the extent that is necessary for measurement;  for all $\rho$ and $x$, $\tr[\E_x \rho]=1 \implies  \ii^L_x(\rho) = \rho$ \cite{Lahti1991}. It follows that if $\rho_x$ is a state prepared by a repeatable $\E$-instrument $\ii$, then a subsequent L\"uders measurement of $\E$ will leave the state $\rho_x$ undisturbed.

\begin{figure}[htbp!]
\begin{center}
\includegraphics[width=0.47\textwidth]{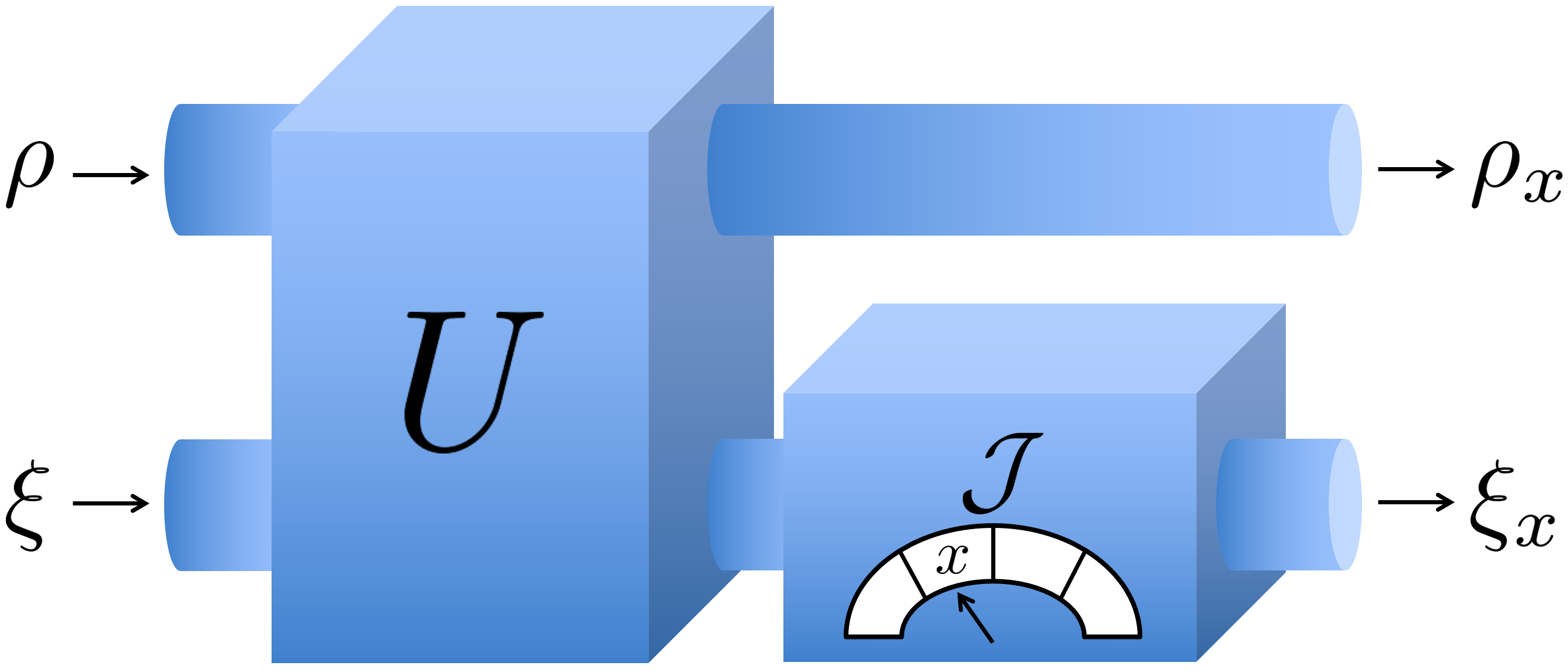}
\vspace*{-0.2cm}
\caption{ The $\E$-instrument $\ii$ is implemented by a  measurement scheme. The system,  prepared in an arbitrary state $\rho$, and the apparatus probe, prepared in the fixed  state $\xi$,  undergo a joint unitary evolution $U$.  Subsequently, the  pointer observable $\Z$ is implemented on the probe by a $\Z$-instrument $\jj$; the probe enters this instrument which registers outcome $x$ with probability $p^\E_\rho (x)   := \tr[\E_x  \rho] = \tr[\ii_x(\rho)]$,  and  transforms the probe to the new state $\xi_x$.   The system is thus transformed to the same state prepared by $\ii$, namely,  $\rho_x$. If $\jj$ is a repeatable $\Z$-instrument then the measurement outcome $x$ is also recorded in the state $\xi_x$; a second measurement of $\Z$ on the probe will produce outcome $x$ with certainty.
}\label{fig:Fig2}
\vspace*{-0.5cm}
\end{center}\
\end{figure}

\subsection{Measurement schemes}
Let us now consider two systems $\s$ and $\aa$, with Hilbert space $\hs$ and $\ha$, respectively. The total system $\s+\aa$ has the Hilbert space $\h:= \hs\otimes \ha$.  We shall consider $\s$ as the system to be measured, and $\aa$ as a probe  that facilitates an indirect measurement of an observable $\E$ of $\s$. An indirect measurement scheme for an $\E$-instrument $\ii$ on $\s$ may be characterised by the tuple  $\mm := (\ha,\xi ,U, \Z)$. Here:  $\xi \in \s(\ha)$ is the initial state of $\aa$;  $U$ is a unitary operator on the composite system $\s+\aa$ which serves to couple the system with the probe; and $\Z := \{\Z_x  : x\in \xx\}$ is a  \emph{pointer observable} of $\aa$,   where  $\xx$  is chosen to be the same value space  as that of the system observable $\E$. The operations of the instrument implemented by $\mm$ may be written as
\begin{align}\label{eq:instrument-dilation}
\ii_x(T) &= \tr\sub{\aa}[(\onesys\otimes \Z_x )U(T \otimes \xi)U^\dagger],
\end{align}
where  $\tr\sub{\aa} : \trc( \h) \to \trc(\hs)$ is the partial trace  over $\aa$, defined by the relation   $\tr[B \, \tr\sub{\aa}[T]] = \tr[(B \otimes \oneapp) T]$ for all $B \in \lo(\hs)$ and $T \in \trc(\h)$. Every $\E$-instrument  $\ii$ admits a \emph{normal} measurement scheme $\mm$, where $\xi$ is a pure state and $\Z$ is a sharp observable  \cite{Ozawa1984}. However, we shall  consider the more general case where  $\xi$ can be a mixed state, and $\Z$ can be unsharp. As with the relationship between instruments and observables, while a measurement scheme $\mm$ corresponds with a unique $\E$-instrument $\ii$, an instrument admits many different measurement schemes. This reflects the fact that one may construct different physical devices, all of which measure the same observable. A schematic of a  measurement scheme  is provided in \fig{fig:Fig2}.

We note that the instrument $\ii$  is independent of how   $\Z$ is measured.  Consider an arbitrary $\Z$-instrument  $\jj$,  and define the identity channel on $\s$ as $\idchsys: \trc(\hs) \to \trc(\hs),  T \mapsto T$. We may therefore define the operation $\idchsys\otimes \jj_x$ as satisfying $\idchsys \otimes \jj_x(T_1 \otimes T_2) = T_1 \otimes \jj_x(T_2)$ for all $T_1 \in \trc(\hs)$ and $T_2 \in \trc(\ha)$. Such an operation can be extended to all of $\trc(\h)$ by linearity.   Let us now define the operations  $\Phi_x : \trc(\hs) \to \trc(\hs)$ as
\begin{align*}
\Phi_x(T) := \tr\sub{\aa} [\idchsys\otimes \jj_x (U T \otimes \xi U^\dagger)],
\end{align*}
 to hold for all $x\in \xx$ and  $T \in \trc(\hs)$. But  for all $x\in \xx$,  $T \in \trc(\hs)$, and $B \in \lo(\hs)$, we have the following:
\begin{align}\label{eq:instrument-dilation-2}
\tr[B \, \Phi_x(T)  ] &= \tr[(B \otimes \oneapp) \idchsys\otimes \jj_x ( U T \otimes \xi U^\dagger ) ], \nonumber \\
&  = \tr[ \idchsys^*\otimes \jj_x^*(B \otimes \oneapp) UT \otimes \xi U^\dagger ], \nonumber \\
&  = \tr[(B \otimes \Z_x) U T \otimes \xi U^\dagger ] , \nonumber \\
& = \tr[B \, \ii_x(T)].
\end{align}
Here, we have used the definition of the partial trace in the first line, the definition of the dual in the second line, the relation $\idchsys^*\otimes \jj_x^*(B \otimes \oneapp) = B\otimes \Z_x$ in the third line, and \eq{eq:instrument-dilation} in the final line.  Since the  equality in \eq{eq:instrument-dilation-2} holds for all $B$ and $T$, it follows that $\Phi_x = \ii_x$ for all $x$. That is to say,   $\ii$ is independent of how the state of the probe changes as a result of the measurement process.

\section{Measurement schemes as a model for the measurement process}\label{sec:measurement-scheme}

Ultimately, all measurements must result from a physical interaction between the system being measured and a measurement apparatus.  We  saw in the discussion above that it is possible to model the measurement process, as a physical interaction between a system $\s$ and a probe  $\aa$, via an indirect measurement scheme $\mm:= (\ha, \xi, U, \Z)$. We now wish to consider $\mm$ as a quantum mechanical model for a direct measurement process, where $\aa$ is  interpreted  as a quantum probe of a macroscopic measurement apparatus.  Given that the boundary between the probe and the rest of the apparatus is arbitrary, we may consider the probe as a quantum mechanical representation of the apparatus itself, and will thus refer to $\aa$  as the apparatus for brevity. 

We may decompose the measurement process into three stages: preparation, premeasurement,  and objectification \cite{PeterMittelstaedt2004}. During preparation the system, initially prepared in an arbitrary state $\rho$, is brought in contact with the  apparatus to prepare the joint state $\rho \otimes \xi$. During premeasurement, the composite system is then evolved by the unitary operator $U$, preparing the joint state  $U(\rho \otimes \xi)U^\dagger$. In general, such a state cannot be understood as a classical mixture of states for which the pointer observable $\Z \equiv \{\onesys \otimes \Z_x\}$ has a definite (objective) value $x$---this is the essential content of the quantum measurement, or pointer objectification, problem \cite{Busch-Shimony-1997, Busch1998a}. Consequently, after premeasurement we must objectify the state $U(\rho \otimes \xi)U^\dagger$ with respect to the pointer observable, thereby preparing the state
\begin{align*}
 \sum_{x\in \xx}p^\E_\rho (x)   \sigma_x,
\end{align*}
where $\tr[(\onesys \otimes \Z_x)\sigma_x] = 1$ for all $x$.  Such a state offers an ignorance interpretation as a classical ensemble of states $\{p^\E_\rho(x), \sigma_x\}$ for which the pointer observable takes definite values $x$ with probabilities $p^\E_\rho(x)$.  In order for pointer objectification to be possible,  we must restrict the pointer observable so that all effects $\Z_x$ have at least one eigenvector with eigenvalue 1, so that each $\sigma_x$ only has support in the eigenvalue-1 eigenspace of $\onesys \otimes \Z_x$;  this implies that the set of states $\{\sigma_x : x \in \xx\}$ will be pairwise orthogonal. Note that pointer objectification implies that it is possible for the observer to ``read'' the measurement outcome without further disturbing the state of system-plus-apparatus; if the observer measures $\Z$ by the L\"uders instrument $\jj^L$, with operations $\jj^L_x(\cdot) := \sqrt{\Z_x} (\cdot) \sqrt{\Z_x}$, then we have $\idchsys \otimes \jj^L_x(\sigma_x) = \sigma_x$ for all $x$.  

   While remaining agnostic as to the precise physical process by which objectification occurs---attempts of physically modelling objectification include, for example, einselection by the environment \cite{Zurek2003}, and the spontaneous collapse model of GRW \cite{Ghirardi1986}---we do demand that it be \emph{some} physical process, that is, a completely positive, trace non-increasing map.        As a conceptual tool, we will therefore consider objectification as a ``measurement'' of the pointer observable $\Z$ by a $\Z$-instrument $\jj$. However, such an instrument must be  \emph{repeatable}.

Given an arbitrary $\Z$-instrument $\jj$,  conditional on producing outcome $x$, the compound of system-plus-apparatus will  be prepared in the   state 
 \begin{align}\label{eq:joint-state-obj}
\sigma_x := \frac{1}{p^\E_\rho (x)  } \idchsys \otimes \jj_x (U(\rho \otimes \xi)U^\dagger), 
 \end{align}
where given \eq{eq:instrument-dilation-2}, we have $p^\E_\rho (x) :=  \tr[\E_x \rho] = \tr[\ii_x(\rho)] = \tr[\idchsys \otimes \jj_x (U(\rho \otimes \xi)U^\dagger)]$. The reduced states of system and apparatus are thus
\begin{align}\label{eq:reduced-state-obj}
&\rho_x   = \tr\sub{\aa}[\sigma_x ], &\xi_x  := \tr\sub{\s}[\sigma_x ], 
\end{align}
where $\rho_x$ is precisely the same state given in \eq{eq:conditional-state}, and $\tr\sub{\s} : \trc( \h) \to \trc(\ha)$ is the partial trace over $\s$.   However, $\sigma_x$ will be objectified with respect to the pointer observable only if 
\begin{align*}
\tr[(\onesys \otimes \Z_x) \sigma_x]= \frac{\tr[\onesys \otimes \jj^*_x(\Z_x)  U(\rho \otimes \xi) U^\dagger]}{\tr[\onesys \otimes \Z_x U(\rho \otimes \xi) U^\dagger]} =1.
\end{align*} 
To ensure this for all input states $\rho$ and outcomes $x$, we must have $\jj_x^*(\Z_x) = \Z_x$, which by \eq{eq:repeatable-defn} implies that  $\jj$ must be a repeatable $\Z$-instrument.      If $\jj$ is a repeatable $\Z$-instrument, which we shall assume henceforth, we shall refer to $\sigma_x$ given in \eq{eq:joint-state-obj}  as the  objectified state, and 
\begin{align}\label{eq:joint-state-obj-average}
\overline{\bm{\sigma}}:= \idchsys \otimes \jj_\xx (U(\rho \otimes \xi)U^\dagger) = \sum_{x\in \xx}p^\E_\rho (x)   \sigma_x
\end{align}
 as the average objectified state.

\section{Internal energy, work, and heat}\label{sec:work-heat}
We now consider the change in internal energy, work, and heat resulting from a measurement scheme $\mm:=(\ha,  \xi , U, \Z)$ for an arbitrary discrete observable $\E$ on a system $\s$, and for an arbitrary system state $\rho$. This discussion follows closely the framework of Ref. \cite{Strasberg2019}. We shall assume that the compound system $\s+\aa$, with Hilbert space $\h:= \hs\otimes \ha$, has  the bounded, additive Hamiltonian $H = H\sub{\s}\otimes \oneapp + \onesys \otimes H\sub{\aa}$, where $H\sub{\s} \in \lo(\hs)$ and $H\sub{\aa} \in \lo(\ha)$ are the  Hamiltonians of each individual system. We refer to  $H$ as the \emph{bare} Hamiltonian, which describes the compound  when both the system and apparatus  are fully isolated. We define the internal energy of  $\s+\aa$, for an arbitrary state $\varrho \in \s(\h)$,  as $\tr[H \varrho]$. The internal energy of $\s$ and $\aa$ are similarly defined. 

The premeasurement stage of the measurement process is  implemented by  introducing a time-dependent interaction Hamiltonian $H_I(t) \in \lo(\h)$, which is non-vanishing only during a finite interval $t\in [t_0, t_1]$. Such time-dependence of the Hamiltonian  generates the  unitary $U$ as
\begin{align*}
U = \overleftarrow{\mathscr{T}} \exp\left(-\imag \int_{t_0}^{t_1} dt \, [ H + H_I(t)] \right),
\end{align*}
with $\overleftarrow{\mathscr{T}}$ denoting the time-ordering operator. The introduction of the time-dependent interaction Hamiltonian $H_I(t)$ is understood to be a purely \emph{mechanical} manipulation of the compound system $\s+\aa$, due to an interaction with  an external macroscopic work source.  Therefore  the increase in internal energy of  $\s+\aa$ during premeasurement will entirely be identified as the \emph{premeasurement work} extracted from the macroscopic source \cite{Allahverdyan2005,Allahverdyan2014}, and will read as  
\begin{align}\label{eq:work}
\ww &:= \tr[(U^\dagger H U - H)\rho \otimes  \xi ],\nonumber \\
& \equiv \tr[H (U(\rho \otimes  \xi) U^\dagger - \rho \otimes  \xi )], \nonumber \\
& = \tr[H\sub{\s}(\ii_\xx(\rho) - \rho)] + \tr[H\sub{\aa}(\eta -  \xi )].
\end{align}
In the final line we use the additivity of $H$ and the definition of the partial trace, where  $\ii_\xx(\rho)  = \tr\sub{\aa}[U(\rho \otimes \xi)U^\dagger]$ and $\eta:= \tr\sub{\s}[U(\rho \otimes  \xi) U^\dagger]$.   Note that $\ww$ is not defined as an average over measurement outcomes, such as is the case in the Two-Point energy Measurement (TPM) protocol \cite{Esposito2009, Campisi2011}, and it can be considered as the  ``unmeasured'' work \cite{Deffner2016a}.  

Recall that after objectification, conditional on producing outcome $x$, the compound system is prepared in the joint objectified state $\sigma_x$ as defined in \eq{eq:joint-state-obj}. Consequently,  for each outcome $x$ such that $p^\E_\rho (x) >0$,  the change in internal energy of the compound of system-plus-apparatus for the full measurement process may be quantified as 
\begin{align}\label{eq:increase-energy}
\Delta \e(x) &:= \tr[H(\sigma_x  - \rho \otimes  \xi )]. 
\end{align}
By the additivity of $H$, we have $\Delta \e(x) =  \Delta \e\sub{\s}(x) + \Delta\e\sub{\aa}(x)$, where  $\Delta \e\sub{\s}(x)$ and $\Delta \e\sub{\aa}(x)$ quantify the change in internal energy of the system and apparatus, respectively, given as 
\begin{align}\label{eq:increase-energy-reduced}
\Delta \e \sub{\s}(x)&:= \tr[H\sub{\s}(\rho_x   - \rho)], \nonumber \\
\Delta\e\sub{\aa}(x) &:= \tr[H\sub{\aa} (\xi_x   -  \xi )],
\end{align}
with $\rho_x$ and $\xi_x$  the reduced states of $\sigma_x$ as defined in  \eq{eq:reduced-state-obj}. For any $x$ such that $p^\E_\rho (x) =0$, we define $\Delta \e\sub{\aa}(x) = \Delta \e\sub{\s}(x) = \Delta \e(x) :=0$.  

We may now consider the first law of thermodynamics. For each outcome $x$ such that $p^\E_\rho (x) >0$, we may define the ``heat'' as $\qq(x) := \Delta \e(x) - \ww$, which is easily obtained from \eq{eq:work} and \eq{eq:increase-energy} to be
\begin{align}\label{eq:heat}
\qq(x)  & =  \tr[H(\sigma_x  -  U(\rho \otimes  \xi) U^\dagger)]. 
\end{align}
For any $x$ such that $p^\E_\rho (x)=0$, we define $\qq(x) := 0$.  This heat is due to the transformation of the premeasured state $U(\rho \otimes  \xi) U^\dagger$ to the objectified states $\sigma_x $, i.e., due to objectification. We may therefore refer to $\qq(x)$ as the \emph{objectification heat}. 

Note that the distribution of $\qq(x)$ is guaranteed to be trivial, that is, $\qq(x)=0$ for all $x$, if:  (i) for the input state $\rho$ an outcome $x$ of the system observable $\E$  is guaranteed (with probability 1) to occur at the outset; and (ii) the pointer observable is sharp and pointer objectification is implemented by the L\"uders $\Z$-instrument $\jj^L$ (which is repeatable if and only if $\Z$ is sharp).  First, note that $p^\E_\rho(x)=1$ if and only if $\rho$ only has support in the eigenvalue-1 eigenspace of $\E_x$, which trivially implies that $\rho$ must commute with $\E$. But it can still be the case that $\rho$ commutes with $\E$, but has support in the eigenspaces of more than one effect, in which case no outcome is definite from the outset. Now, given that $\mm$ is a measurement scheme for $\E$,   $p^\E_\rho(x)=1 \implies \tr[(\onesys \otimes \Z_x) U(\rho \otimes \xi)U^\dagger]=1$. But recall that the L\"uders instrument is ideal, i.e., $\tr[\Z_x \varrho]=1 \implies \jj^L_x(\varrho) = \varrho$. It follows that in such a case we have $\idchsys\otimes \jj^L_x(U(\rho \otimes \xi)U^\dagger) =  U(\rho \otimes \xi)U^\dagger$, and hence $\sigma_x = U(\rho \otimes \xi)U^\dagger$, which by \eq{eq:heat} trivially gives $\qq(x) = 0$. If the pointer observable is sharp, and pointer objectification is implemented by the L\"uders instrument, we can conclude that the heat distribution will be non-trivial only if the measurement outcomes of the system observable $\E$ in the input state $\rho$ are indeterminate. But note that  if pointer objectification is implemented by an arbitrary repeatable instrument $\jj$, compatible with a possibly unsharp observable $\Z$, it may still be the case that $p^\E_\rho(x) =1$ but nonetheless we have $\sigma_x \ne U(\rho \otimes \xi)U^\dagger$, and it will be possible to have $\qq(x) \ne 0$.  

As an aside, let us note that $\Delta \e(x)$ defined in \eq{eq:increase-energy} is not fully conditional on the measurement outcome $x$; while the final energy $ \tr[H \sigma_x]$ depends on $x$,  the initial energy $\tr[H \rho \otimes  \xi ]$ does not. In \app{app:conditional-energy} we compare the present approach to that suggested in Ref. \cite{Mohammady2019c}, where the initial energy is also conditioned on the measurement outcome. This method motivates a definition for the conditional work, whereby applying the first law leads to a drastically different interpretation of ``heat'' as a counter-factual quantity.

Finally, upon averaging over all measurement outcomes we obtain the following:
\begin{align}\label{eq:average-heat-energy}
\avg{\Delta \e} &:= \sum_{x\in \xx} p^\E_\rho (x)   \Delta \e(x) = \tr[H(\overline{\bm{\sigma}} - \rho\otimes \xi)], \nonumber \\
\avg{\qq} &:= \sum_{x\in \xx} p^\E_\rho (x)   \qq(x) = \tr[H (\overline{\bm{\sigma}} - U(\rho\otimes \xi) U^\dagger)], 
\end{align} 
where $\overline{\bm{\sigma}}$ is the average objectified state defined in \eq{eq:joint-state-obj-average}. Combining these with the work, we thus obtain the average first law as 
\begin{align}\label{eq:average-first-law-1}
\avg{\Delta \e}  = \ww + \avg{\qq}. 
\end{align}
Note that  $\tr\sub{\aa}[\overline{\bm{\sigma}}] = \tr\sub{\aa}[U(\rho\otimes \xi) U^\dagger] = \ii_\xx(\rho)$ always holds. As such, the average heat will only depend on the apparatus degrees of freedom, and can be equivalently expressed as $\avg{\qq} = \tr[H\sub{\aa} (\jj_\xx(\eta) - \eta)]$, where  $\jj_\xx(\eta)\equiv \sum_x p^\E_\rho (x) \xi_x$.   

\section{Necessity of the Yanase condition}\label{sec:Yanase}

In the previous section, we identified work with the premeasurement stage of the measurement process---the work exchanged with an external source as a result of inducing time-dependence on the compound system's Hamiltonian so as to generate the unitary evolution $U$. However, once the premeasurement stage is complete  the compound of system-plus-apparatus is once again governed by the time-independent bare Hamiltonian $H = H\sub{\s}\otimes \oneapp + \onesys \otimes H\sub{\aa}$, and no work is exchanged with an external source thereafter. Indeed, this is a crucial assumption for the energetic changes during objectification to be fully identified as heat.   

Since after premeasurement the compound system $\s+\aa$ has the time-independent bare Hamiltonian $H$, it follows that  $\s+\aa$ will be governed by ``time-translation'' symmetry. Here the (compact, abelian) symmetry group is $G= \re$, with the strongly continuous unitary representation in $\h:= \hs\otimes \ha$ generated by the Hamiltonian as $V: G \ni g \mapsto V(g) :=  e^{-\imag g H}$. By additivity of $H$, we may also write $V(g) = V\sub{\s}(g) \otimes V\sub{\aa}(g)$, where $V\sub{\s}(g):= e^{-\imag g H\sub{\s}}$ and $V\sub{\aa}(g):= e^{-\imag g H\sub{\aa}}$. Below, we shall provide two arguments as to why time-translation symmetry demands that the pointer observable must commute with the Hamiltonian: (i) The time at which objectification takes place should not make a physically observable difference; and (ii) the record of the measurement outcomes produced by objectification should be time-independent.  We refer to the commutation of the pointer observable with the  Hamiltonian as the \emph{Yanase condition} \cite{Yanase1961, Ozawa2002}, first introduced in the context of the Wigner-Araki-Yanase (WAY) theorem \cite{E.Wigner1952,  Busch2010, Araki1960}.   

Let us first consider (i). This can be justified heuristically by considering that the observer has no way of  \emph{knowing} precisely at what time after premeasurement objectification takes place. Specifically, the observer cannot distinguish between the following states of affairs: (\emph{a}) The pointer observable is immediately objectified after premeasurement, and then the compound system evolves for some time $g$; and (\emph{b}) the compound system evolves for some time $g$ after premeasurement, and then the pointer observable is objectified. Such indiscernibility implies that the operations of the $\Z$-instrument $\jj$  must be time-translation \emph{covariant}:
\begin{align*}
&\idchsys \otimes \jj_x (V(g) \cdot V(g)^\dagger) = V(g) \idchsys \otimes \jj_x ( \cdot ) V(g)^\dagger,\\
& \qquad \implies \jj_x(V\sub{\aa}(g) \cdot V\sub{\aa}(g)^\dagger) = V\sub{\aa}(g) \jj_x(\cdot) V\sub{\aa}(g)^\dagger,
\end{align*} 
to hold for all $x\in \xx$ and  $g\in G$, where the second line follows from the additivity of $H$.  Noting that $\Z_x = \jj_x^*(\oneapp)$ by definition, and that covariance in the Schr\"odinger picture is equivalent to covariance in the Heisenberg picture, we have for all $x\in \xx$ and $g \in G$ the following:
\begin{align*}
\Z_x &=   \jj_x^*( V\sub{\aa}(g)^\dagger \oneapp  V\sub{\aa}(g)) =  V\sub{\aa}(g)^\dagger \ii^*_x(\oneapp)  V\sub{\aa}(g) , \nonumber \\
&= V\sub{\aa}(g)^\dagger \Z_x  V\sub{\aa}(g).
\end{align*}
 In other words, the Heisenberg evolved pointer observable $\Z(g) := \{\Z_x(g) \equiv V\sub{\aa}(g)^\dagger \Z_x  V\sub{\aa}(g) : x\in \xx\}$ must equal the pointer observable $\Z$ for all $g$;  time-translation covariance of $\jj$ implies that $\Z$ must be time-translation \emph{invariant}. Indeed, as argued by Loveridge et al in  Ref. \cite{Loveridge2017a}, the only obserable quantities of a system governed by a symmetry group $G$ are those that are invariant under its action.   By computing the differential of both sides of the equality $\Z_x(g) = \Z_x$  with respect to $g$, we obtain $ [\Z_x(g), H\sub{\aa}] = \zero$.  By evaluating this commutator as $g\to 0$, we see that $[\Z_x, H\sub{\aa}]=\zero$ must hold for all $x$, which we shall denote  by the short-hand $ [\Z, H\sub{\aa}]=\zero$.

Now let us consider (ii). Even if we are to assume that pointer objectification can occur with a non-invariant pointer observable, the Yanase condition can be argued for \emph{a fortiori} on the basis of the stability of the measurement outcomes. Let us assume that the compound system is objectified with respect to an arbitrary pointer observable $\Z$ immediately after premeasurement,  producing outcome $x$ and thus preparing the objectified state $\sigma_x$, defined in \eq{eq:joint-state-obj}, which only has support in the eigenvalue-1 eigenspace of $\Z_x$.  Now, assume that  the external observer chooses to read the measurement outcome by measuring $\Z$ at some time $g$ after objectification. It follows that  the  observer will detect outcome $x$ with the probability 
\begin{align}\label{eq:stability-outcome}
 \tr[\onesys \otimes  \Z_x (g)  \sigma_x ] = \frac{\tr[\onesys \otimes \jj_x^*(\Z_x (g)) U(\rho \otimes \xi) U^\dagger]}{\tr[\onesys \otimes \Z_x U(\rho \otimes \xi) U^\dagger]}.
\end{align}
   The record of the measurement outcome is stable (or time-independent) if and only if for all $\rho \in \s(\hs)$, $x\in \xx$,  and  $g \in G$,  \eq{eq:stability-outcome} equals $1$.  It is easy to see that this will be satisfied only if  $\Z_x(g) =  \Z_x$  for all $x$ and $g$, so that by repeatability of $\jj$ we obtain $\jj_x^*(\Z_x(g)) = \jj_x^*(\Z_x) =  \Z_x$. Once again,   the Yanase condition must hold.

\section{Uncertainty of the objectification heat}\label{sec:heat-uncertainty}

In the previous sections we defined the heat $\qq$ that results as the compound of the system to be measured, and the measurement apparatus, is objectified with respect to the pointer observable. But using symmetry principles and the requirement that the objectified values be stable across time, we argued that the pointer observable must commute with the Hamiltonian, that is, the Yanase condition must be fulfilled. Now we wish to consider what implications the Yanase condition will have for the statistics of the objectification heat. 

First, let us note that if the Hamiltonian of the apparatus is a fixed point of the $\Z$-channel $\jj_\xx^*$, i.e., $\jj_\xx^*(H\sub{\aa})= H\sub{\aa}$, then the average objectification heat will vanish for all input states $\rho$. In \app{app:zero-average-heat} we show that in the case where the pointer observable is sharp and satisfies the Yanase condition, and either (i) objectification is implemented by the L\"uders instrument $\jj^L$, or (ii) the Hamiltonian can be written as $H\sub{\aa} = \sum_x \epsilon_x \Z_x$,  then $\jj_\xx^*(H\sub{\aa})= H\sub{\aa}$ will always hold. However, we provide a simple counter example where even if the Yanase condition is fulfilled, it still holds that $\jj_\xx^*(H\sub{\aa}) \ne H\sub{\aa}$, and so it will be possible for some input states to have $\avg{\qq}\ne 0$. Of course, the average heat is not the only quantity of interest. The fluctuations, or \emph{uncertainty},  of the heat is also informative. As we show below,   the Yanase condition guarantees that the uncertainty of the objectification heat is fully classical.

 The uncertainty of the  objectification heat $\qq$ is defined as the variance $\var{\qq} := \avg{\qq^2} - \avg{\qq}^2$ which, as shown in \app{app:heat-variance},  can always be written as
\begin{align}\label{eq:variance-heat-1}
\var{\qq} = \mathrm{V}(H,  \overline{\bm{\sigma}}) - \sum_{x\in \xx} p^\E_\rho (x)    \mathrm{V}(H,  \sigma_x ).
\end{align}
Here $\sigma_x $ and $\overline{\bm{\sigma}}:= \sum_x p^\E_\rho (x) \sigma_x$ are the  states defined in \eq{eq:joint-state-obj} and \eq{eq:joint-state-obj-average}, respectively,  while for any self-adjoint $A \in \lo(\h)$ and $\varrho \in \s(\h)$,  the variance of $A$ in $\varrho$ is defined as $ \mathrm{V}(A, \varrho) := \tr[A^2 \varrho] - \tr[A \varrho]^2$.  

To disambiguate the classical and quantum contributions to $\var{\qq}$, let us first note that  the variance $\mathrm{V}(A, \varrho)$ can be split into a classical and quantum component as
\begin{align}\label{eq:variance-skew-info-relation}
 \mathrm{V}(A, \varrho) &= \qv(A,\varrho) + \cv(A, \varrho), \nonumber \\
\qv(A,\varrho) &:= \tr[A^2 \varrho] - \tr[A \varrho^\alpha A \varrho^{1-\alpha}], \nonumber \\
\cv(A,\varrho) &:= \tr[A \varrho^\alpha A \varrho^{1-\alpha}] - \tr[A \varrho]^2,
\end{align}
where $\alpha \in (0,1)$ and, for a state with spectral decomposition $\varrho = \sum_i p_i P_i$, we define $\varrho^\alpha := \sum_i p_i^\alpha P_i$.  $\qv(A,\varrho)$ is the Wigner-Yanase-Dyson skew information \cite{Wigner1963, Lieb1973} which: (i) is non-negative and bounded by the variance $0 \leqslant \qv(A, \varrho) \leqslant \mathrm{V}(A, \varrho)$; (ii) reduces to $\mathrm{V}(A, \varrho)$ if $\varrho$ is a pure state and vanishes if $[A, \varrho] = \zero$;  and (iii) is convex under classical mixing, i.e., $\sum_i p_i \qv(A,\varrho_i) \geqslant \qv(A,\sum_i p_i \varrho_i)$. Conditions (i)-(iii) satisfy the definition for a measure of quantum uncertainty proposed in Ref. \cite{Luo2005}, and hence  $\qv(A,\varrho)$ can be understood as quantifying the quantum uncertainty of $A$ in $\varrho$. On the other hand, $\cv(A,\varrho)$ may be interpreted as quantifying  the remaining classical uncertainty of $A$ in $\varrho$, and it: (i) is non-negative and bounded by the variance $0 \leqslant \cv(A, \varrho) \leqslant \mathrm{V}(A, \varrho)$; (ii)  reduces to  $\mathrm{V}(A, \varrho)$ if $[A, \varrho] = \zero$ and vanishes if $\varrho$ is a pure state;  and (iii) is concave under classical mixing, i.e., $\sum_i p_i \cv(A,\varrho_i) \leqslant \cv(A,\sum_i p_i \varrho_i)$.

The uncertainty of energy in the average objectified state, and the objectified states, can be split into a quantum and classical component defined in  \eq{eq:variance-skew-info-relation}  as $\mathrm{V}(H, \overline{\bm{\sigma}}) = \qv(H, \overline{\bm{\sigma}}) + \cv(H, \overline{\bm{\sigma}})$ and $\mathrm{V}(H, \sigma_x) = \qv(H, \sigma_x) + \cv(H, \sigma_x)$, respectively. Therefore,  we may now write the uncertainty of heat shown in \eq{eq:variance-heat-1} as 
\begin{align}\label{eq:variance-heat-quantum-classical}
\var{\qq} & =  \Delta \cv - \Delta \qv,
\end{align}
where we define 
\begin{align}\label{eq:quantum-classical-cont-variance-heat}
\Delta \qv &:= \sum_{x\in \xx} p^\E_\rho (x)   \qv(H,\sigma_x ) - \qv(H,\overline{\bm{\sigma}})  \geqslant 0 , \nonumber \\ 
\Delta \cv &:= \cv(H,\overline{\bm{\sigma}}) - \sum_{x\in \xx} p^\E_\rho (x)   \cv(H,\sigma_x ) \geqslant 0 ,
\end{align}
with positivity  ensured by Lieb's concavity theorem \cite{Lieb1973}. $\Delta \qv $ quantifies the decrease in quantum uncertainty of energy when the objectified states $\sigma_x$ are classically mixed. Conversely, $\Delta \cv$ quantifies the increase in classical uncertainty of energy when the objectified states are classically mixed.  Note that $\var{\qq}\geqslant 0$ implies that   $\Delta \cv \geqslant \Delta \qv $ always holds.  

The expressions presented thus far hold for all pointer observables $\Z$, and all implementations $\jj$.  Now, let us assume that the $\Z$-instrument $\jj$ is repeatable, which is necessary for pointer objectification.  Repeatability of $\jj$ implies  that  $\{\sigma_x :x \in \xx \}$ are pairwise orthogonal, with each $\sigma_x$ only having support in the eigenvalue-1 eigenspace of the effects $\onesys \otimes \Z_x $.  Repeatability thus implies that  $\overline{\bm{\sigma}}$ and $\{\sigma_x : x \in \xx\}$ have a common set of spectral projections, and so for any $\alpha$  we  may write    $\overline{\bm{\sigma}}^\alpha = \sum_x p^\E_\rho (x)^\alpha \sigma_x ^\alpha$.   Consequently, by using \eq{eq:variance-skew-info-relation} and \eq{eq:quantum-classical-cont-variance-heat}, and noting that $\sum_x p^\E_\rho (x) \tr[H^2 \sigma_x] = \tr[H^2 \overline{\bm{\sigma}}]$,  we may rewrite  $\Delta \qv$ as 
\begin{align}\label{eq:quantum-contribution-variance-heat}
\Delta \qv  &= \tr[H \overline{\bm{\sigma}}^\alpha H\overline{\bm{\sigma}}^{1-\alpha}] \nonumber \\
& \qquad - \sum_{x\in \xx} p^\E_\rho (x)   \tr[H \sigma_x ^\alpha H \sigma_x ^{1-\alpha}], \nonumber \\
&= \sum_{x \ne y} p^\E_\rho (x)^\alpha p^\E_\rho (y)^{1-\alpha} \tr[H\sigma_x^\alpha H\sigma_y^{1-\alpha}]. 
\end{align}
Now let us also assume that $\Z$ satisfies the  Yanase condition $[\Z, H\sub{\aa}] =\zero$ which,  by additivity of $H$, is equivalent to $[\onesys\otimes \Z, H]=\zero$.  We recall that repeatability of $\jj$  implies  the   identities $\sigma_x^\alpha =  \sigma_x^\alpha (\onesys \otimes P_x ) = (\onesys \otimes P_x ) \sigma_x^\alpha$ for all $x$ and $\alpha$, where $P_x$ denotes   the projection onto the eigenvalue-1 eigenspace of $\Z_x$. Consequently,  we will have for all $x\ne y$ the following:
\begin{align*}
\sigma_x^\alpha H\, \sigma_y^{1-\alpha} &= \sigma_x^\alpha(\onesys\otimes P_x)  H (\onesys \otimes P_y)  \sigma_y^{1-\alpha},  \nonumber \\
&=\sigma_x^\alpha H (\onesys \otimes P_x  P_y)   \sigma_y^{1-\alpha}, \nonumber \\
& = \zero.
\end{align*}
Here,   in the second line we have used  the Yanase condition  which, since $P_x$ is a spectral projection of $\Z_x$,  implies that $[\onesys \otimes P_x, H]=\zero$,  and in the final line we have used the fact that $P_x P_y = \zero$  if $x\ne y$. We see by  \eq{eq:quantum-contribution-variance-heat}  that $\Delta \qv = 0$,  and so \eq{eq:variance-heat-quantum-classical} reduces to  
\begin{align}\label{eq:variance-heat-classical}
\var{\qq} & =  \Delta \cv.
\end{align}
The uncertainty in objectification heat is entirely identified with $\Delta \cv$, which quantifies the increase in classical uncertainty of energy by classically mixing the objectified states $\sigma_x$ to prepare the mixture $\overline{\bm{\sigma}}$. While the quantum uncertainty of energy in the states $\overline{\bm{\sigma}}$ and $\{\sigma_x : x \in \xx\}$ need not vanish individually, such uncertainty will play no role in the magnitude of $\var{\qq} $. As such, we may interpret the uncertainty of the objectification heat $\qq$ as being entirely classical.

As a remark, we note that the same  arguments as above will apply, \emph{mutatis mutandis}, for the uncertainty of $\Delta \e\sub{\aa}$ as defined in \eq{eq:increase-energy-reduced}, that is, the uncertainty of the  change in internal energy of the measurement apparatus. Repeatability of the $\Z$-instrument $\jj$ will mean that the conditional apparatus states $\{\xi_x : x\in \xx\}$ will be pairwise-orthogonal, with each $\xi_x$ only having support in the eigenvalue-1 eigenspace of $\Z_x$.  The Yanase condition will thus imply   that the quantum contribution to $\var{\Delta \e\sub{\aa}} := \avg{\Delta \e\sub{\aa}^2} - \avg{\Delta \e\sub{\aa}}^2$ will be strictly zero.

\subsection{Measurement, heat, and  information transfer}

To interpret the classicality of the heat uncertainty in information theoretic terms, let us conceive the objectification process as resulting from a measurement of the pointer observable by a fictitious agent which we shall refer to as a \emph{Daimon}---from the greek $\delta \alpha \iota \mu \omega \nu$, the root of which means ``to divide''. At first, the Daimon measures the pointer observable $\Z$ by a repeatable $\Z$-instrument, thereby  ``encoding'' the information regarding the measurement outcome $x$ in the objectified state $\sigma_x$ defined in \eq{eq:joint-state-obj}. Such information can be perfectly transmitted to the observer, since a second measurement of the pointer observable by the observer will recover outcome $x$ on the state $\sigma_x$ with certainty. However, the Daimon may also choose to encode ``time information'' in the objectified states, by allowing them to evolve according to their isolated Hamiltonian evolution, with such time evolution being dependent on the outcome $x$ observed. As we shall see below,  the Yanase condition will ensure perfect transfer of both types of information to the observer, with the perfect information transfer regarding time being concomitant with the classical uncertainty of heat.

In order to describe the process of information transfer, let us first assign a Hilbert space $\h\sub{\dd}$ to the Daimon's memory $\dd$, and  denote by $\{\ket{x}\}$ an orthonormal basis of $\h\sub{\dd}$, with each $\ket{x}$ indicating that the Daimon has observed outcome $x$ of the pointer observable $\Z$ (and hence of the system observable $\E$). We shall also assign a Hamiltonian $H\sub{\dd}$ to the Daimon's memory, and since such memory should be time-independent, then $\{\ket{x}\}$ are also eigenstates of $H\sub{\dd}$. Since the memory of the Daimon is perfectly correlated with the measurement outcomes, then conditional on observing outcome $x$, the compound system $\dd+ \s + \aa$ will be prepared in the state $|x\>\<x| \otimes \sigma_x$.  Conditional on observing outcome $x$, the Daimon may then allow the compound system $\s+\aa$ to evolve for time $g_x$, where $g_x = g_y \iff x=y$. The Daimon  thus prepares the joint state $|x\>\<x| \otimes \sigma_x(g_x)$, where $\sigma_x(g_x) :=  e^{-\imag g_x H} (\sigma_x) e^{\imag g_x H}$, with $H = H\sub{\s}\otimes \oneapp + \onesys \otimes H\sub{\aa}$ the additive Hamiltonian of the compound system $\s+\aa$. The average joint state of the Daimon's memory and the compound of system-plus-apparatus can thus be represented as 
\begin{align}\label{eq:daimon-system-apparatus-state}
\varrho := \sum_x p^\E_\rho(x) |x\>\<x| \otimes \sigma_x(g_x), 
\end{align}
where    $p^\E_\rho(x):= \tr[\E_x \rho] = \tr[(\onesys \otimes \Z_x )U(\rho \otimes \xi)U^\dagger]$.  The process of information transfer from the Daimon to the observer is described by the partial trace channel $\tr\sub{\dd} : \trc(\h\sub{\dd} \otimes \h) \to \trc(\h)$, so that the observer receives the compound of system-plus-apparatus in the state $\tr\sub{\dd}[\varrho] = \overline{\bm{\sigma}}(\overline{g}) := \sum_x p^\E_\rho(x) \sigma_x(g_x)$.

For any state $\rho$, the von Neumann entropy is defined as $S(\rho) := -\tr[\rho  \ln(\rho)]$. We may quantify the information content of $\varrho$ defined in \eq{eq:daimon-system-apparatus-state} pertaining to the measurement outcomes $x$ by the von Neumann entropy $S(\varrho)$.    Note that for any collection of pairwise orthogonal states $\{\rho_i\}$, we always have $S(\sum_i p_i \rho_i) = \mathscr{H} + \sum_i p_i S(\rho_i)$, where $\mathscr{H} \equiv \mathscr{H}(\{p_i\}) := -\sum_i p_i \ln(p_i)$ is the Shannon entropy of the probability distribution $\{p_i\}$. Since the states $|x\>\<x| \otimes \sigma_x(g_x)$ are pairwise orthogonal, while $|x\>\<x|$ are pure states and the von Neumann entropy is invariant under unitary evolution, we have $S(\varrho) = \sum_{x} p^\E_\rho(x) S(\sigma_x) + \mathscr{H}$, with $\mathscr{H} \equiv \mathscr{H}(\{p^\E_\rho(x)\})$ the Shannon entropy of the measurement probability distribution. It follows that 
\begin{align}\label{eq:von-Neumann-entropy-loss}
S(\overline{\bm{\sigma}}(\overline{g})) - S(\varrho) = \mathscr{X} - \mathscr{H} \leqslant 0,
\end{align}
where $\mathscr{X} := S(\overline{\bm{\sigma}}(\overline{g})) - \sum_{x} p^\E_\rho(x) S(\sigma_x) \leqslant \mathscr{H} $ is the Holevo information \cite{Holevo1973}, which quantifies the maximum amount of classical information pertaining to the random variable $x$ that can be transmitted given the ensemble $\{p^\E_\rho(x), \sigma_x(g_x)\}$. By the inequality in \eq{eq:von-Neumann-entropy-loss}, we see that the information received by the observer cannot be greater than that obtained by the Daimon. 

But recall that the Daimon implements $\Z$ by a repeatable instrument $\jj$, which is required by objectification, and that $\Z$ satisfies the Yanase condition $[ \Z, H\sub{\aa}] = [\onesys \otimes \Z, H]=\zero$, which is required for the objectified values to be stable.  Repeatability guarantees that $\{\sigma_x\}$ will be pairwise orthogonal, with each $\sigma_x$ only having support in the  eigenvalue-1 eigenspace of $\onesys \otimes \Z_x$. Since the Yanase condition implies that $[\onesys \otimes P_x, H]=\zero$ for all $x$, where $P_x$ is the projection onto the eigenvalue-1 eigenspace of $\Z_x$,  it follows that $\{\sigma_x(g_x)\}$ are also pairwise orthogonal, since 
\begin{align*}
\sigma_x(g_x)&=  e^{-\imag g_x H}(\onesys \otimes P_x)\sigma_x (\onesys \otimes P_x)e^{\imag g_x H}, \nonumber \\
& = (\onesys \otimes P_x) e^{-\imag g_x H} \sigma_x e^{\imag g_x H} (\onesys \otimes P_x), \nonumber \\
& = (\onesys \otimes P_x)  \sigma_x(g_x) (\onesys \otimes P_x).
\end{align*} 
Note that orthogonality of $\{\sigma_x(g_x)\}$ is guaranteed by repeatability alone if $g_x = g$ for all $x$, since  $\sigma_x(g) \sigma_y(g) = e^{-\imag g_x H}\sigma_x e^{\imag g_x H}e^{-\imag g_y H} \sigma_y e^{\imag g_y H} = \zero $ if $x\ne y$ and $g_x = g_y$. But we assume that $g_x$ are distinct, and so $e^{\imag g_x H}e^{-\imag g_y H} = e^{\imag (g_x - g_y) H} \ne \one$.  The orthogonality of the states $\{\sigma_x(g_x)\}$ implies that $S(\overline{\bm{\sigma}}(\overline{g})) = \sum_{x} p^\E_\rho(x) S(\sigma_x) + \mathscr{H}$, and hence $\mathscr{X} = \mathscr{H}$,   so that the upper bound of \eq{eq:von-Neumann-entropy-loss} is saturated;  the observer's measurement of the pointer observable $\Z$ on the apparatus will recover outcomes $x$ by the probability distribution $p^\E_\rho(x)$, and so  none of the classical information regarding the measurement outcomes is lost as such information is transmitted to the observer. But if the collection of states $\{\sigma_x(g_x)\}$ are not pairwise orthogonal---implying that either $\jj$ is not repeatable, meaning that the pointer observable was not objectified,  or the Yanase condition is violated, meaning that the objectified values are not stable---it is known that $ \mathscr{X} \leqslant \kappa \mathscr{H}$, where $\kappa < 1$ quantifies the maximum trace-distance between the states $\sigma_x(g_x)$ \cite{Audenaert2014}. In such a case \eq{eq:von-Neumann-entropy-loss} becomes a strict inequality, indicating a non-vanishing loss of information.

Now let us turn to the other type of information transfer, that is, information regarding time. For a system governed by a Hamiltonian $H$, the ``asymmetry'' of a state $\rho$ with reference to $H$ may be quantified by the Wigner-Yanase-Dyson skew information $\qv(H, \rho)$ defined in \eq{eq:variance-skew-info-relation}. If $\rho$ commutes with $H$, so that $\qv(H, \rho) = 0$, then $e^{-\imag g H} \rho e^{\imag g H} = \rho$ for all $g$, and so $\rho$ does not contain any information regarding $g$. On the other hand, if $\rho$ does not commute with $H$ so that $\qv(H, \rho)$ is large, then the orbit of states $\{e^{-\imag g H} \rho e^{\imag g H} : g\}$ will also be large, in which case $\rho$ serves as a better encoding of  $g$ \cite{Marvian2014}. We may therefore quantify the information regarding time (the random variables $\{g_x\}$), encoded in the state $\varrho$, by the skew information as $\qv(H\sub{\mathrm{tot}}, \varrho)$, where   $H\sub{\mathrm{tot}}:= H\sub{\dd} \otimes \one + \one\sub{\dd} \otimes H$ is the total, additive Hamiltonian of the compound system $\dd + \s + \aa$. Since the skew information  never increases under channels that are covariant with respect to time-translation symmetry, and  the partial trace channel is always time-translation covariant \cite{Takagi2018}, we thus have $\qv(H\sub{\mathrm{tot}}, \varrho) \geqslant \qv(H, \overline{\bm{\sigma}}(\overline{g}))$. But given that $\ket{x}$ are orthogonal eigenstates of $H\sub{\dd}$, as shown in  \app{app:skew-info-identity} we always have  $\qv(H\sub{\mathrm{tot}}, \varrho) = \sum_x p^\E_\rho(x) \qv(H, \sigma_x(g_x)) = \sum_x p^\E_\rho(x) \qv(H, \sigma_x)$, where the second equality follows from the fact that the skew information is invariant under unitary evolutions generated by $H$. It follows that
\begin{align}\label{eq:skew-info-loss}
&\qv(H, \overline{\bm{\sigma}}(\overline{g})) - \qv(H\sub{\mathrm{tot}}, \varrho) \nonumber \\
& \qquad = \qv(H, \overline{\bm{\sigma}}(\overline{g})) - \sum_x p^\E_\rho(x) \qv(H, \sigma_x)  \leqslant 0.
\end{align}
Once again, by the arguments preceding \eq{eq:variance-heat-classical}, repeatability and the Yanase condition will guarantee that $\qv(H, \overline{\bm{\sigma}}(\overline{g})) = \sum_x p^\E_\rho(x) \qv(H, \sigma_x)$ holds, in which case the upper bound  of \eq{eq:skew-info-loss} is saturated, meaning that none of the information regarding time has been lost as the objectified states are received by the observer. But this is precisely what it means for the uncertainty of the objectification heat to be entirely classical, as the quantum contribution to such uncertainty is exactly identified with the loss of time-information, or asymmetry, as the objectified states $\sigma_x$ are classically mixed.

\section{Case study: Normal measurement scheme for the L\"uders instrument}\label{sec:Luders}

We shall now provide a simple example to illustrate the general observations made above. Consider an arbitrary discrete observable $\E$ of the system $\s$,    with $\mm:=(\ha, \ket{\varphi}, U, \Z)$ a normal measurement scheme for the L\"uders $\E$-instrument $\ii^L$, with the operations $\ii^L_x(\cdot) = \sqrt{\E_x} (\cdot)\sqrt{\E_x} $. Here, the apparatus is initially prepared in the pure state $\xi = |\varphi\>\<\varphi|$, and the pointer observable is sharp. As before we assume that the compound system has the total, additive Hamiltonian $H = H\sub{\s} \otimes \oneapp + \onesys \otimes H\sub{\aa}$, and that the pointer observable satisfies the Yanase condition $[\Z, H\sub{\aa}]=\zero$. 

Without loss of generality, we may characterise the action of $U$ as
\begin{align*}
U : \ket{\psi} \otimes \ket{\varphi} \mapsto \sum_{x\in \xx} \sqrt{\E_x} \ket{\psi} \otimes \ket{\phi_x},
\end{align*}
to hold for all $\ket{\psi} \in \hs$. Here, $\ket{\phi_x}\in \ha$ are pairwise orthogonal unit vectors satisfying the relation $\Z_x \ket{\phi_y} = \delta_{x,y} \ket{\phi_x}$, that is, $\ket{\phi_x}$  are eigenvalue-1 eigenstates of the projection operators $\Z_x$. Note that the Yanase condition  implies that $\ket{\phi_x}$ are also eigenstates of $H\sub{\aa}$.  It is simple to verify that after premeasurement, for an arbitrary input state $\rho$ we prepare the joint state
\begin{align}\label{eq:Luders-premeas-state}
U(\rho \otimes |\varphi\>\<\varphi| )U^\dagger = \sum_{x,y \in \xx} \sqrt{\E_x} \rho \sqrt{\E_y} \otimes |\phi_x\>\<\phi_y|. 
\end{align}
The reduced states of system and apparatus, after premeasurement, are 
\begin{align}\label{eq:Luders-premeas-red-state}
\tr\sub{\aa}[U(\rho \otimes |\varphi\>\<\varphi| )U^\dagger] &=  \sum_{x\in \xx} \sqrt{\E_x} \rho \sqrt{\E_x} = \ii^L_\xx(\rho), \nonumber \\
\tr\sub{\s}[U(\rho \otimes |\varphi\>\<\varphi| )U^\dagger] & = \sum_{x,y \in \xx} \tr[\sqrt{\E_y} \sqrt{\E_x} \rho] |\phi_x\>\<\phi_y| =: \eta.
\end{align}

For simplicity, let us also model objectification by the L\"uders instrument compatible with $\Z$, which is repeatable as $\Z$ is sharp. For all outcomes $x$ that obtain with a non-vanishing probability, we thus have the objectified states
\begin{align}\label{eq:Luders-obj-state}
\sigma_x &:= \frac{\idchsys \otimes \jj^L_x (U(\rho \otimes |\varphi\>\<\varphi| )U^\dagger)}{\tr[(\onesys \otimes \Z_x)U(\rho \otimes |\varphi\>\<\varphi| )U^\dagger]}, \nonumber \\
& = \rho_x \otimes |\phi_x\>\<\phi_x|,
\end{align}
where we define $\rho_x:= \ii^L_x(\rho)/p^\E_\rho(x) \equiv \sqrt{\E_x} \rho \sqrt{\E_x}/\tr[\E_x \rho] $, with the reduced states 
\begin{align}\label{eq:Luders-obj-red-state}
\tr\sub{\aa}[\sigma_x] &= \rho_x, \nonumber \\
\tr\sub{\s}[\sigma_x] &=  |\phi_x\>\<\phi_x|.
\end{align}
Similarly, the average objectified state is
\begin{align}\label{eq:Luders-avg-obj-state}
\overline{\bm{\sigma}} &:= \idchsys \otimes \jj^L_\xx(U(\rho \otimes |\varphi\>\<\varphi|)U^\dagger), \nonumber \\
&= \sum_{x\in \xx} p^\E_\rho(x) \rho_x \otimes |\phi_x\>\<\phi_x|,
\end{align}
with the reduced states
\begin{align}\label{eq:Luders-avg-obj-red-state}
\tr\sub{\aa}[\overline{\bm{\sigma}}] &= \ii^L_\xx(\rho), \nonumber \\
\tr\sub{\s}[\overline{\bm{\sigma}}] &= \sum_{x\in \xx} p^\E_\rho(x) |\phi_x\>\<\phi_x|= \jj^L_\xx(\eta).
\end{align}

By \eq{eq:work}, \eq{eq:Luders-premeas-state}, and \eq{eq:Luders-premeas-red-state} the work done during premeasurement will read
\begin{align*}
\ww &:= \tr[(U^* H U - H)\rho \otimes |\varphi\>\<\varphi|], \nonumber \\
& = \tr[H\sub{\s} (\ii^L_\xx(\rho) - \rho)] + \tr[H\sub{\aa} (\jj^L_\xx(\eta) - |\varphi\>\<\varphi|)],
\end{align*}
where in the final line we use the fact that since $\Z$ commutes with $H\sub{\aa}$, then $\tr[H\sub{\aa} \eta] = \tr[{\jj^L_\xx}^*(H\sub{\aa}) \eta] = \tr[H\sub{\aa} \jj^L_\xx(\eta)]$. If $\E$ commutes with $H\sub{\s}$, we also have  $\tr[H\sub{\s} (\ii^L_\xx(\rho) - \rho)]  = \tr[({\ii^L_\xx}^*(H\sub{\s}) - H\sub{\s}) \rho] = 0$, and  in such a case the work will be entirely determined by the change in average internal energy of the apparatus. 

By \eq{eq:heat}, and Eqs. \eqref{eq:Luders-obj-state}-\eqref{eq:Luders-avg-obj-red-state} the objectification heat will read
\begin{align*}
\qq(x) &= \tr[H(\sigma_x - U(\rho \otimes |\varphi\>\<\varphi|)U^\dagger)], \nonumber \\
& = \tr[H \sigma_x] - \tr[H \idchsys \otimes \jj^L_\xx(U(\rho \otimes |\varphi\>\<\varphi|)U^\dagger))], \nonumber \\
& = \tr[H(\sigma_x - \overline{\bm{\sigma}})], \nonumber \\
& = \tr[H\sub{\s} (\rho_x  - \ii^L_\xx(\rho))] + \tr[H\sub{\aa} (|\phi_x\>\<\phi_x|  - \jj^L_\xx(\eta))].
\end{align*}
In the second line we use the fact that $\Z$ satisfies the Yanase condition, and that objectification is implemented by the L\"uders instrument, which gives $H\sub{\aa} = {\jj^L_\xx}^*(H\sub{\aa})  \implies H = {\idchsys \otimes \jj^L_\xx}^*(H)$. In other words, the heat is simply the difference between the expected energy of the objectifed state $\sigma_x$ and the average objectified state $\overline{\bm{\sigma}}$. Assuming that $H\sub{\s}$ and $H\sub{\aa}$ have a fully non-degenerate spectrum, then it is simple to see that the heat distribution will be trivial, i.e., $\qq(x) = 0$ for all $x$, if and only if the measurement outcome is certain from the outset. If only one outcome $x$ obtains with probability 1, then $\overline{\bm{\sigma}} = \sigma_x$, and so $\qq(x) =0$. But if more than one outcome $x$ obtains, then $\rho_x  \ne \ii^L_\xx(\rho)$ and $|\phi_x\>\<\phi_x|  \ne \jj^L_\xx(\eta)$, and non-degeneracy of the  spectrum of $H\sub{\s}$ and $H\sub{\aa}$ implies that $\qq(x)$ must be non-vanishing for at least one outcome $x$ that is observed. 
 
It is trivial to see  that the average objectification heat always vanishes, i.e., $\avg{\qq}=0$ for all $\rho$ and $\E$.  As for the quantum contribution to the uncertainty in $\qq$, by  \eq{eq:quantum-classical-cont-variance-heat} we have

\begin{align*}
\Delta \qv &= \sum_{x\in \xx} p^\E_\rho (x)   \qv(H,\sigma_x ) - \qv(H,\overline{\bm{\sigma}}).
\end{align*}
That the above quantity vanishes follows immediately from \eq{eq:Luders-obj-state} and \eq{eq:Luders-avg-obj-state}, and  the fact that $|\phi_x\>$ are orthogonal eigenstates of $H\sub{\aa}$. As shown in \app{app:skew-info-identity}, this implies that $\qv(H,\overline{\bm{\sigma}}) = \sum_{x\in \xx} p^\E_\rho (x)   \qv(H\sub{\s},\rho_x ) $ and $\qv(H,\sigma_x ) = \qv(H\sub{\s},\rho_x )$.

\section{Conclusions}
We have considered the physical implementation of a general discrete observable as a \emph{measurement scheme}, which is decomposed into three stages: (i) preparation, where the system to be measured is brought in contact with a measurement apparatus initially prepared in a fixed state; (ii) premeasurement, involving  a unitary  interaction between the system  and   apparatus; (iii) and  pointer objectification, whereby the compound of system-plus-apparatus is transformed to a state for which the pointer observable assumes definite values, which can then be ``read'' by the observer without causing further disturbance.     We identified the work with premeasurement, and the heat with  objectification. In order for the apparatus to serve as a stable record for the measurement outcomes of the system observable, we demanded that the pointer observable commute with the Hamiltonian, i.e., satisfy the Yanase condition. We showed that the Yanase condition ensures that the  uncertainty of the heat resulting from objectification  is entirely classical, and identified such classicality as being concomitant with perfect information transfer to the observer.

\begin{acknowledgments} This project has received funding from the European Union’s Horizon 2020 research and innovation program under the Marie Skłodowska-Curie grant agreement No 801505, as well as  from  the Slovak Academy of Sciences   under MoRePro project OPEQ (19MRP0027).
\end{acknowledgments}


\bibliographystyle{apsrev4-2}
\bibliography{References}

\begin{thebibliography}{79}%
\makeatletter
\providecommand \@ifxundefined [1]{%
 \@ifx{#1\undefined}
}%
\providecommand \@ifnum [1]{%
 \ifnum #1\expandafter \@firstoftwo
 \else \expandafter \@secondoftwo
 \fi
}%
\providecommand \@ifx [1]{%
 \ifx #1\expandafter \@firstoftwo
 \else \expandafter \@secondoftwo
 \fi
}%
\providecommand \natexlab [1]{#1}%
\providecommand \enquote  [1]{``#1''}%
\providecommand \bibnamefont  [1]{#1}%
\providecommand \bibfnamefont [1]{#1}%
\providecommand \citenamefont [1]{#1}%
\providecommand \href@noop [0]{\@secondoftwo}%
\providecommand \href [0]{\begingroup \@sanitize@url \@href}%
\providecommand \@href[1]{\@@startlink{#1}\@@href}%
\providecommand \@@href[1]{\endgroup#1\@@endlink}%
\providecommand \@sanitize@url [0]{\catcode `\\12\catcode `\$12\catcode
  `\&12\catcode `\#12\catcode `\^12\catcode `\_12\catcode `\%12\relax}%
\providecommand \@@startlink[1]{}%
\providecommand \@@endlink[0]{}%
\providecommand \url  [0]{\begingroup\@sanitize@url \@url }%
\providecommand \@url [1]{\endgroup\@href {#1}{\urlprefix }}%
\providecommand \urlprefix  [0]{URL }%
\providecommand \Eprint [0]{\href }%
\providecommand \doibase [0]{https://doi.org/}%
\providecommand \selectlanguage [0]{\@gobble}%
\providecommand \bibinfo  [0]{\@secondoftwo}%
\providecommand \bibfield  [0]{\@secondoftwo}%
\providecommand \translation [1]{[#1]}%
\providecommand \BibitemOpen [0]{}%
\providecommand \bibitemStop [0]{}%
\providecommand \bibitemNoStop [0]{.\EOS\space}%
\providecommand \EOS [0]{\spacefactor3000\relax}%
\providecommand \BibitemShut  [1]{\csname bibitem#1\endcsname}%
\let\auto@bib@innerbib\@empty
\bibitem [{\citenamefont {Funo}\ \emph {et~al.}(2013)\citenamefont {Funo},
  \citenamefont {Watanabe},\ and\ \citenamefont {Ueda}}]{Funo2013}%
  \BibitemOpen
  \bibfield  {author} {\bibinfo {author} {\bibfnamefont {K.}~\bibnamefont
  {Funo}}, \bibinfo {author} {\bibfnamefont {Y.}~\bibnamefont {Watanabe}},\
  and\ \bibinfo {author} {\bibfnamefont {M.}~\bibnamefont {Ueda}},\ }\href
  {https://doi.org/10.1103/PhysRevE.88.052121} {\bibfield  {journal} {\bibinfo
  {journal} {Phys. Rev. E}\ }\textbf {\bibinfo {volume} {88}},\ \bibinfo
  {pages} {052121} (\bibinfo {year} {2013})}\BibitemShut {NoStop}%
\bibitem [{\citenamefont {Roncaglia}\ \emph {et~al.}(2014)\citenamefont
  {Roncaglia}, \citenamefont {Cerisola},\ and\ \citenamefont
  {Paz}}]{Roncaglia2014}%
  \BibitemOpen
  \bibfield  {author} {\bibinfo {author} {\bibfnamefont {A.~J.}\ \bibnamefont
  {Roncaglia}}, \bibinfo {author} {\bibfnamefont {F.}~\bibnamefont
  {Cerisola}},\ and\ \bibinfo {author} {\bibfnamefont {J.~P.}\ \bibnamefont
  {Paz}},\ }\href {https://doi.org/10.1103/PhysRevLett.113.250601} {\bibfield
  {journal} {\bibinfo  {journal} {Phys. Rev. Lett.}\ }\textbf {\bibinfo
  {volume} {113}},\ \bibinfo {pages} {250601} (\bibinfo {year}
  {2014})}\BibitemShut {NoStop}%
\bibitem [{\citenamefont {Watanabe}\ \emph {et~al.}(2014)\citenamefont
  {Watanabe}, \citenamefont {Venkatesh},\ and\ \citenamefont
  {Talkner}}]{Watanabe2014}%
  \BibitemOpen
  \bibfield  {author} {\bibinfo {author} {\bibfnamefont {G.}~\bibnamefont
  {Watanabe}}, \bibinfo {author} {\bibfnamefont {B.~P.}\ \bibnamefont
  {Venkatesh}},\ and\ \bibinfo {author} {\bibfnamefont {P.}~\bibnamefont
  {Talkner}},\ }\href {https://doi.org/10.1103/PhysRevE.89.052116} {\bibfield
  {journal} {\bibinfo  {journal} {Phys. Rev. E}\ }\textbf {\bibinfo {volume}
  {89}},\ \bibinfo {pages} {052116} (\bibinfo {year} {2014})}\BibitemShut
  {NoStop}%
\bibitem [{\citenamefont {Perarnau-Llobet}\ \emph {et~al.}(2017)\citenamefont
  {Perarnau-Llobet}, \citenamefont {B{\"{a}}umer}, \citenamefont
  {Hovhannisyan}, \citenamefont {Huber},\ and\ \citenamefont
  {Acin}}]{Perarnau-Llobet2016a}%
  \BibitemOpen
  \bibfield  {author} {\bibinfo {author} {\bibfnamefont {M.}~\bibnamefont
  {Perarnau-Llobet}}, \bibinfo {author} {\bibfnamefont {E.}~\bibnamefont
  {B{\"{a}}umer}}, \bibinfo {author} {\bibfnamefont {K.~V.}\ \bibnamefont
  {Hovhannisyan}}, \bibinfo {author} {\bibfnamefont {M.}~\bibnamefont
  {Huber}},\ and\ \bibinfo {author} {\bibfnamefont {A.}~\bibnamefont {Acin}},\
  }\href {https://doi.org/10.1103/PhysRevLett.118.070601} {\bibfield  {journal}
  {\bibinfo  {journal} {Phys. Rev. Lett.}\ }\textbf {\bibinfo {volume} {118}},\
  \bibinfo {pages} {070601} (\bibinfo {year} {2017})}\BibitemShut {NoStop}%
\bibitem [{\citenamefont {Morikuni}\ \emph {et~al.}(2017)\citenamefont
  {Morikuni}, \citenamefont {Tajima},\ and\ \citenamefont
  {Hatano}}]{Morikuni2017}%
  \BibitemOpen
  \bibfield  {author} {\bibinfo {author} {\bibfnamefont {Y.}~\bibnamefont
  {Morikuni}}, \bibinfo {author} {\bibfnamefont {H.}~\bibnamefont {Tajima}},\
  and\ \bibinfo {author} {\bibfnamefont {N.}~\bibnamefont {Hatano}},\ }\href
  {https://doi.org/10.1103/PhysRevE.95.032147} {\bibfield  {journal} {\bibinfo
  {journal} {Phys. Rev. E}\ }\textbf {\bibinfo {volume} {95}},\ \bibinfo
  {pages} {032147} (\bibinfo {year} {2017})}\BibitemShut {NoStop}%
\bibitem [{\citenamefont {Elouard}\ \emph
  {et~al.}(2017{\natexlab{a}})\citenamefont {Elouard}, \citenamefont
  {Bernardes}, \citenamefont {Carvalho}, \citenamefont {Santos},\ and\
  \citenamefont {Auff{\`{e}}ves}}]{Elouard2017b}%
  \BibitemOpen
  \bibfield  {author} {\bibinfo {author} {\bibfnamefont {C.}~\bibnamefont
  {Elouard}}, \bibinfo {author} {\bibfnamefont {N.~K.}\ \bibnamefont
  {Bernardes}}, \bibinfo {author} {\bibfnamefont {A.~R.~R.}\ \bibnamefont
  {Carvalho}}, \bibinfo {author} {\bibfnamefont {M.~F.}\ \bibnamefont
  {Santos}},\ and\ \bibinfo {author} {\bibfnamefont {A.}~\bibnamefont
  {Auff{\`{e}}ves}},\ }\href {https://doi.org/10.1088/1367-2630/aa7fa2}
  {\bibfield  {journal} {\bibinfo  {journal} {New J. Phys.}\ }\textbf {\bibinfo
  {volume} {19}},\ \bibinfo {pages} {103011} (\bibinfo {year}
  {2017}{\natexlab{a}})}\BibitemShut {NoStop}%
\bibitem [{\citenamefont {{De Chiara}}\ \emph {et~al.}(2018)\citenamefont {{De
  Chiara}}, \citenamefont {Solinas}, \citenamefont {Cerisola},\ and\
  \citenamefont {Roncaglia}}]{DeChiara2018}%
  \BibitemOpen
  \bibfield  {author} {\bibinfo {author} {\bibfnamefont {G.}~\bibnamefont {{De
  Chiara}}}, \bibinfo {author} {\bibfnamefont {P.}~\bibnamefont {Solinas}},
  \bibinfo {author} {\bibfnamefont {F.}~\bibnamefont {Cerisola}},\ and\
  \bibinfo {author} {\bibfnamefont {A.~J.}\ \bibnamefont {Roncaglia}},\ }in\
  \href {https://doi.org/10.1007/978-3-319-99046-0_14} {\emph {\bibinfo
  {booktitle} {Thermodyn. Quantum Regime. Fundam. Theor. Phys.}}}\ (\bibinfo
  {publisher} {Springer},\ \bibinfo {year} {2018})\ pp.\ \bibinfo {pages}
  {337--362}\BibitemShut {NoStop}%
\bibitem [{\citenamefont {Sone}\ \emph {et~al.}(2020)\citenamefont {Sone},
  \citenamefont {Liu},\ and\ \citenamefont {Cappellaro}}]{Sone2020}%
  \BibitemOpen
  \bibfield  {author} {\bibinfo {author} {\bibfnamefont {A.}~\bibnamefont
  {Sone}}, \bibinfo {author} {\bibfnamefont {Y.-X.}\ \bibnamefont {Liu}},\ and\
  \bibinfo {author} {\bibfnamefont {P.}~\bibnamefont {Cappellaro}},\ }\href
  {https://doi.org/10.1103/PhysRevLett.125.060602} {\bibfield  {journal}
  {\bibinfo  {journal} {Phys. Rev. Lett.}\ }\textbf {\bibinfo {volume} {125}},\
  \bibinfo {pages} {060602} (\bibinfo {year} {2020})}\BibitemShut {NoStop}%
\bibitem [{\citenamefont {Mohammady}(2021)}]{Mohammady2020}%
  \BibitemOpen
  \bibfield  {author} {\bibinfo {author} {\bibfnamefont {M.~H.}\ \bibnamefont
  {Mohammady}},\ }\href {https://doi.org/10.1103/PhysRevA.103.042214}
  {\bibfield  {journal} {\bibinfo  {journal} {Phys. Rev. A}\ }\textbf {\bibinfo
  {volume} {103}},\ \bibinfo {pages} {042214} (\bibinfo {year}
  {2021})}\BibitemShut {NoStop}%
\bibitem [{\citenamefont {Hovhannisyan}\ and\ \citenamefont
  {Imparato}(2021)}]{Hovhannisyan2021}%
  \BibitemOpen
  \bibfield  {author} {\bibinfo {author} {\bibfnamefont {K.~V.}\ \bibnamefont
  {Hovhannisyan}}\ and\ \bibinfo {author} {\bibfnamefont {A.}~\bibnamefont
  {Imparato}},\ }\href {http://arxiv.org/abs/2104.09364} {\  (\bibinfo {year}
  {2021})},\ \Eprint {https://arxiv.org/abs/2104.09364} {arXiv:2104.09364}
  \BibitemShut {NoStop}%
\bibitem [{\citenamefont {Jacobs}(2009)}]{Jacobs2009}%
  \BibitemOpen
  \bibfield  {author} {\bibinfo {author} {\bibfnamefont {K.}~\bibnamefont
  {Jacobs}},\ }\href {https://doi.org/10.1103/PhysRevA.80.012322} {\bibfield
  {journal} {\bibinfo  {journal} {Phys. Rev. A}\ }\textbf {\bibinfo {volume}
  {80}},\ \bibinfo {pages} {012322} (\bibinfo {year} {2009})}\BibitemShut
  {NoStop}%
\bibitem [{\citenamefont {Shiraishi}\ \emph {et~al.}(2015)\citenamefont
  {Shiraishi}, \citenamefont {Ito}, \citenamefont {Kawaguchi},\ and\
  \citenamefont {Sagawa}}]{Shiraishi2015a}%
  \BibitemOpen
  \bibfield  {author} {\bibinfo {author} {\bibfnamefont {N.}~\bibnamefont
  {Shiraishi}}, \bibinfo {author} {\bibfnamefont {S.}~\bibnamefont {Ito}},
  \bibinfo {author} {\bibfnamefont {K.}~\bibnamefont {Kawaguchi}},\ and\
  \bibinfo {author} {\bibfnamefont {T.}~\bibnamefont {Sagawa}},\ }\href
  {https://doi.org/10.1088/1367-2630/17/4/045012} {\bibfield  {journal}
  {\bibinfo  {journal} {New J. Phys.}\ }\textbf {\bibinfo {volume} {17}},\
  \bibinfo {pages} {045012} (\bibinfo {year} {2015})}\BibitemShut {NoStop}%
\bibitem [{\citenamefont {Hayashi}\ and\ \citenamefont
  {Tajima}(2017)}]{Hayashi2017}%
  \BibitemOpen
  \bibfield  {author} {\bibinfo {author} {\bibfnamefont {M.}~\bibnamefont
  {Hayashi}}\ and\ \bibinfo {author} {\bibfnamefont {H.}~\bibnamefont
  {Tajima}},\ }\href {https://doi.org/10.1103/PhysRevA.95.032132} {\bibfield
  {journal} {\bibinfo  {journal} {Phys. Rev. A}\ }\textbf {\bibinfo {volume}
  {95}},\ \bibinfo {pages} {032132} (\bibinfo {year} {2017})}\BibitemShut
  {NoStop}%
\bibitem [{\citenamefont {Mohammady}\ and\ \citenamefont
  {Anders}(2017)}]{Mohammady2017}%
  \BibitemOpen
  \bibfield  {author} {\bibinfo {author} {\bibfnamefont {M.~H.}\ \bibnamefont
  {Mohammady}}\ and\ \bibinfo {author} {\bibfnamefont {J.}~\bibnamefont
  {Anders}},\ }\href {https://doi.org/10.1088/1367-2630/aa8ba1} {\bibfield
  {journal} {\bibinfo  {journal} {New J. Phys.}\ }\textbf {\bibinfo {volume}
  {19}},\ \bibinfo {pages} {113026} (\bibinfo {year} {2017})}\BibitemShut
  {NoStop}%
\bibitem [{\citenamefont {Elouard}\ \emph
  {et~al.}(2017{\natexlab{b}})\citenamefont {Elouard}, \citenamefont
  {Herrera-Mart{\'{i}}}, \citenamefont {Huard},\ and\ \citenamefont
  {Auff{\`{e}}ves}}]{Elouard2017a}%
  \BibitemOpen
  \bibfield  {author} {\bibinfo {author} {\bibfnamefont {C.}~\bibnamefont
  {Elouard}}, \bibinfo {author} {\bibfnamefont {D.}~\bibnamefont
  {Herrera-Mart{\'{i}}}}, \bibinfo {author} {\bibfnamefont {B.}~\bibnamefont
  {Huard}},\ and\ \bibinfo {author} {\bibfnamefont {A.}~\bibnamefont
  {Auff{\`{e}}ves}},\ }\href {https://doi.org/10.1103/PhysRevLett.118.260603}
  {\bibfield  {journal} {\bibinfo  {journal} {Phys. Rev. Lett.}\ }\textbf
  {\bibinfo {volume} {118}},\ \bibinfo {pages} {260603} (\bibinfo {year}
  {2017}{\natexlab{b}})}\BibitemShut {NoStop}%
\bibitem [{\citenamefont {Elouard}\ and\ \citenamefont
  {Jordan}(2018)}]{Elouard2018b}%
  \BibitemOpen
  \bibfield  {author} {\bibinfo {author} {\bibfnamefont {C.}~\bibnamefont
  {Elouard}}\ and\ \bibinfo {author} {\bibfnamefont {A.~N.}\ \bibnamefont
  {Jordan}},\ }\href {https://doi.org/10.1103/PhysRevLett.120.260601}
  {\bibfield  {journal} {\bibinfo  {journal} {Phys. Rev. Lett.}\ }\textbf
  {\bibinfo {volume} {120}},\ \bibinfo {pages} {260601} (\bibinfo {year}
  {2018})}\BibitemShut {NoStop}%
\bibitem [{\citenamefont {Manzano}\ \emph {et~al.}(2018)\citenamefont
  {Manzano}, \citenamefont {Plastina},\ and\ \citenamefont
  {Zambrini}}]{Manzano2018}%
  \BibitemOpen
  \bibfield  {author} {\bibinfo {author} {\bibfnamefont {G.}~\bibnamefont
  {Manzano}}, \bibinfo {author} {\bibfnamefont {F.}~\bibnamefont {Plastina}},\
  and\ \bibinfo {author} {\bibfnamefont {R.}~\bibnamefont {Zambrini}},\ }\href
  {https://doi.org/10.1103/PhysRevLett.121.120602} {\bibfield  {journal}
  {\bibinfo  {journal} {Phys. Rev. Lett.}\ }\textbf {\bibinfo {volume} {121}},\
  \bibinfo {pages} {120602} (\bibinfo {year} {2018})}\BibitemShut {NoStop}%
\bibitem [{\citenamefont {Buffoni}\ \emph {et~al.}(2019)\citenamefont
  {Buffoni}, \citenamefont {Solfanelli}, \citenamefont {Verrucchi},
  \citenamefont {Cuccoli},\ and\ \citenamefont {Campisi}}]{Buffoni2018}%
  \BibitemOpen
  \bibfield  {author} {\bibinfo {author} {\bibfnamefont {L.}~\bibnamefont
  {Buffoni}}, \bibinfo {author} {\bibfnamefont {A.}~\bibnamefont {Solfanelli}},
  \bibinfo {author} {\bibfnamefont {P.}~\bibnamefont {Verrucchi}}, \bibinfo
  {author} {\bibfnamefont {A.}~\bibnamefont {Cuccoli}},\ and\ \bibinfo {author}
  {\bibfnamefont {M.}~\bibnamefont {Campisi}},\ }\href
  {https://doi.org/10.1103/PhysRevLett.122.070603} {\bibfield  {journal}
  {\bibinfo  {journal} {Phys. Rev. Lett.}\ }\textbf {\bibinfo {volume} {122}},\
  \bibinfo {pages} {070603} (\bibinfo {year} {2019})}\BibitemShut {NoStop}%
\bibitem [{\citenamefont {Solfanelli}\ \emph {et~al.}(2019)\citenamefont
  {Solfanelli}, \citenamefont {Buffoni}, \citenamefont {Cuccoli},\ and\
  \citenamefont {Campisi}}]{Solfanelli2019}%
  \BibitemOpen
  \bibfield  {author} {\bibinfo {author} {\bibfnamefont {A.}~\bibnamefont
  {Solfanelli}}, \bibinfo {author} {\bibfnamefont {L.}~\bibnamefont {Buffoni}},
  \bibinfo {author} {\bibfnamefont {A.}~\bibnamefont {Cuccoli}},\ and\ \bibinfo
  {author} {\bibfnamefont {M.}~\bibnamefont {Campisi}},\ }\href
  {https://doi.org/10.1088/1742-5468/ab3721} {\bibfield  {journal} {\bibinfo
  {journal} {J. Stat. Mech. Theory Exp.}\ }\textbf {\bibinfo {volume} {2019}},\
  \bibinfo {pages} {094003} (\bibinfo {year} {2019})}\BibitemShut {NoStop}%
\bibitem [{\citenamefont {Purves}\ and\ \citenamefont
  {Short}(2021)}]{Purves-2020}%
  \BibitemOpen
  \bibfield  {author} {\bibinfo {author} {\bibfnamefont {T.}~\bibnamefont
  {Purves}}\ and\ \bibinfo {author} {\bibfnamefont {A.~J.}\ \bibnamefont
  {Short}},\ }\href {https://doi.org/10.1103/PhysRevE.104.014111} {\bibfield
  {journal} {\bibinfo  {journal} {Phys. Rev. E}\ }\textbf {\bibinfo {volume}
  {104}},\ \bibinfo {pages} {014111} (\bibinfo {year} {2021})}\BibitemShut
  {NoStop}%
\bibitem [{\citenamefont {Bresque}\ \emph {et~al.}(2021)\citenamefont
  {Bresque}, \citenamefont {Camati}, \citenamefont {Rogers}, \citenamefont
  {Murch}, \citenamefont {Jordan},\ and\ \citenamefont
  {Auff{\`{e}}ves}}]{Bresque2020}%
  \BibitemOpen
  \bibfield  {author} {\bibinfo {author} {\bibfnamefont {L.}~\bibnamefont
  {Bresque}}, \bibinfo {author} {\bibfnamefont {P.~A.}\ \bibnamefont {Camati}},
  \bibinfo {author} {\bibfnamefont {S.}~\bibnamefont {Rogers}}, \bibinfo
  {author} {\bibfnamefont {K.}~\bibnamefont {Murch}}, \bibinfo {author}
  {\bibfnamefont {A.~N.}\ \bibnamefont {Jordan}},\ and\ \bibinfo {author}
  {\bibfnamefont {A.}~\bibnamefont {Auff{\`{e}}ves}},\ }\href
  {https://doi.org/10.1103/PhysRevLett.126.120605} {\bibfield  {journal}
  {\bibinfo  {journal} {Phys. Rev. Lett.}\ }\textbf {\bibinfo {volume} {126}},\
  \bibinfo {pages} {120605} (\bibinfo {year} {2021})}\BibitemShut {NoStop}%
\bibitem [{\citenamefont {Zurek}(2003{\natexlab{a}})}]{Zurek2003a}%
  \BibitemOpen
  \bibfield  {author} {\bibinfo {author} {\bibfnamefont {W.~H.}\ \bibnamefont
  {Zurek}},\ }\href {https://arxiv.org/abs/quant-ph/0301076} {\  (\bibinfo
  {year} {2003}{\natexlab{a}})},\ \Eprint {https://arxiv.org/abs/quant-ph/0301076}
  {arXiv:0301076 [quant-ph]} \BibitemShut {NoStop}%
\bibitem [{\citenamefont {Miyadera}(2011)}]{Miyadera2011d}%
  \BibitemOpen
  \bibfield  {author} {\bibinfo {author} {\bibfnamefont {T.}~\bibnamefont
  {Miyadera}},\ }\href {https://doi.org/10.1103/PhysRevA.83.052119} {\bibfield
  {journal} {\bibinfo  {journal} {Phys. Rev. A}\ }\textbf {\bibinfo {volume}
  {83}},\ \bibinfo {pages} {052119} (\bibinfo {year} {2011})}\BibitemShut
  {NoStop}%
\bibitem [{\citenamefont {Navascu{\'{e}}s}\ and\ \citenamefont
  {Popescu}(2014)}]{Navascues2014a}%
  \BibitemOpen
  \bibfield  {author} {\bibinfo {author} {\bibfnamefont {M.}~\bibnamefont
  {Navascu{\'{e}}s}}\ and\ \bibinfo {author} {\bibfnamefont {S.}~\bibnamefont
  {Popescu}},\ }\href {https://doi.org/10.1103/PhysRevLett.112.140502}
  {\bibfield  {journal} {\bibinfo  {journal} {Phys. Rev. Lett.}\ }\textbf
  {\bibinfo {volume} {112}},\ \bibinfo {pages} {140502} (\bibinfo {year}
  {2014})}\BibitemShut {NoStop}%
\bibitem [{\citenamefont {Miyadera}(2016)}]{Miyadera2015a}%
  \BibitemOpen
  \bibfield  {author} {\bibinfo {author} {\bibfnamefont {T.}~\bibnamefont
  {Miyadera}},\ }\href {https://doi.org/10.1007/s10701-016-0027-6} {\bibfield
  {journal} {\bibinfo  {journal} {Found. Phys.}\ }\textbf {\bibinfo {volume}
  {46}},\ \bibinfo {pages} {1522} (\bibinfo {year} {2016})}\BibitemShut
  {NoStop}%
\bibitem [{\citenamefont {Allahverdyan}\ \emph {et~al.}(2017)\citenamefont
  {Allahverdyan}, \citenamefont {Balian},\ and\ \citenamefont
  {Nieuwenhuizen}}]{Allahverdyan2013a}%
  \BibitemOpen
  \bibfield  {author} {\bibinfo {author} {\bibfnamefont {A.~E.}\ \bibnamefont
  {Allahverdyan}}, \bibinfo {author} {\bibfnamefont {R.}~\bibnamefont
  {Balian}},\ and\ \bibinfo {author} {\bibfnamefont {T.~M.}\ \bibnamefont
  {Nieuwenhuizen}},\ }\href {https://doi.org/10.1016/j.aop.2016.11.001}
  {\bibfield  {journal} {\bibinfo  {journal} {Ann. Phys. (N. Y).}\ }\textbf
  {\bibinfo {volume} {376}},\ \bibinfo {pages} {324} (\bibinfo {year}
  {2017})}\BibitemShut {NoStop}%
\bibitem [{\citenamefont {Konishi}(2018)}]{Konishi2016a}%
  \BibitemOpen
  \bibfield  {author} {\bibinfo {author} {\bibfnamefont {E.}~\bibnamefont
  {Konishi}},\ }\href {https://doi.org/10.1088/1742-5468/aac13f} {\bibfield
  {journal} {\bibinfo  {journal} {J. Stat. Mech. Theory Exp.}\ }\textbf
  {\bibinfo {volume} {2018}},\ \bibinfo {pages} {063403} (\bibinfo {year}
  {2018})}\BibitemShut {NoStop}%
\bibitem [{\citenamefont {Mancino}\ \emph {et~al.}(2018)\citenamefont
  {Mancino}, \citenamefont {Sbroscia}, \citenamefont {Roccia}, \citenamefont
  {Gianani}, \citenamefont {Somma}, \citenamefont {Mataloni}, \citenamefont
  {Paternostro},\ and\ \citenamefont {Barbieri}}]{Mancino2017}%
  \BibitemOpen
  \bibfield  {author} {\bibinfo {author} {\bibfnamefont {L.}~\bibnamefont
  {Mancino}}, \bibinfo {author} {\bibfnamefont {M.}~\bibnamefont {Sbroscia}},
  \bibinfo {author} {\bibfnamefont {E.}~\bibnamefont {Roccia}}, \bibinfo
  {author} {\bibfnamefont {I.}~\bibnamefont {Gianani}}, \bibinfo {author}
  {\bibfnamefont {F.}~\bibnamefont {Somma}}, \bibinfo {author} {\bibfnamefont
  {P.}~\bibnamefont {Mataloni}}, \bibinfo {author} {\bibfnamefont
  {M.}~\bibnamefont {Paternostro}},\ and\ \bibinfo {author} {\bibfnamefont
  {M.}~\bibnamefont {Barbieri}},\ }\href
  {https://doi.org/10.1038/s41534-018-0069-z} {\bibfield  {journal} {\bibinfo
  {journal} {npj Quantum Inf.}\ }\textbf {\bibinfo {volume} {4}},\ \bibinfo
  {pages} {20} (\bibinfo {year} {2018})}\BibitemShut {NoStop}%
\bibitem [{\citenamefont {Benoist}\ \emph {et~al.}(2018)\citenamefont
  {Benoist}, \citenamefont {Jak{\v{s}}i{\'{c}}}, \citenamefont {Pautrat},\ and\
  \citenamefont {Pillet}}]{Benoist2017}%
  \BibitemOpen
  \bibfield  {author} {\bibinfo {author} {\bibfnamefont {T.}~\bibnamefont
  {Benoist}}, \bibinfo {author} {\bibfnamefont {V.}~\bibnamefont
  {Jak{\v{s}}i{\'{c}}}}, \bibinfo {author} {\bibfnamefont {Y.}~\bibnamefont
  {Pautrat}},\ and\ \bibinfo {author} {\bibfnamefont {C.-A.}\ \bibnamefont
  {Pillet}},\ }\href {https://doi.org/10.1007/s00220-017-2947-1} {\bibfield
  {journal} {\bibinfo  {journal} {Commun. Math. Phys.}\ }\textbf {\bibinfo
  {volume} {357}},\ \bibinfo {pages} {77} (\bibinfo {year} {2018})}\BibitemShut
  {NoStop}%
\bibitem [{\citenamefont {Guryanova}\ \emph {et~al.}(2020)\citenamefont
  {Guryanova}, \citenamefont {Friis},\ and\ \citenamefont
  {Huber}}]{Guryanova2018}%
  \BibitemOpen
  \bibfield  {author} {\bibinfo {author} {\bibfnamefont {Y.}~\bibnamefont
  {Guryanova}}, \bibinfo {author} {\bibfnamefont {N.}~\bibnamefont {Friis}},\
  and\ \bibinfo {author} {\bibfnamefont {M.}~\bibnamefont {Huber}},\ }\href
  {https://doi.org/10.22331/q-2020-01-13-222} {\bibfield  {journal} {\bibinfo
  {journal} {Quantum}\ }\textbf {\bibinfo {volume} {4}},\ \bibinfo {pages}
  {222} (\bibinfo {year} {2020})}\BibitemShut {NoStop}%
\bibitem [{\citenamefont {Benoist}\ \emph {et~al.}(2021)\citenamefont
  {Benoist}, \citenamefont {Cuneo}, \citenamefont {Jak{\v{s}}i{\'{c}}},\ and\
  \citenamefont {Pillet}}]{Benoist2020}%
  \BibitemOpen
  \bibfield  {author} {\bibinfo {author} {\bibfnamefont {T.}~\bibnamefont
  {Benoist}}, \bibinfo {author} {\bibfnamefont {N.}~\bibnamefont {Cuneo}},
  \bibinfo {author} {\bibfnamefont {V.}~\bibnamefont {Jak{\v{s}}i{\'{c}}}},\
  and\ \bibinfo {author} {\bibfnamefont {C.~A.}\ \bibnamefont {Pillet}},\
  }\href {https://doi.org/10.1007/s10955-021-02725-1} {\bibfield  {journal}
  {\bibinfo  {journal} {J. Stat. Phys.}\ }\textbf {\bibinfo {volume} {182}},\
  \bibinfo {pages} {44} (\bibinfo {year} {2021})}\BibitemShut {NoStop}%
\bibitem [{\citenamefont {Naikoo}\ \emph {et~al.}(2021)\citenamefont {Naikoo},
  \citenamefont {Banerjee}, \citenamefont {Pan},\ and\ \citenamefont
  {Ghosh}}]{Naikoo2021}%
  \BibitemOpen
  \bibfield  {author} {\bibinfo {author} {\bibfnamefont {J.}~\bibnamefont
  {Naikoo}}, \bibinfo {author} {\bibfnamefont {S.}~\bibnamefont {Banerjee}},
  \bibinfo {author} {\bibfnamefont {A.~K.}\ \bibnamefont {Pan}},\ and\ \bibinfo
  {author} {\bibfnamefont {S.}~\bibnamefont {Ghosh}},\ }\href
  {http://arxiv.org/abs/2103.00974} {\  (\bibinfo {year} {2021})},\ \Eprint
  {https://arxiv.org/abs/2103.00974} {arXiv:2103.00974} \BibitemShut {NoStop}%
\bibitem [{\citenamefont {Landi}\ \emph {et~al.}(2021)\citenamefont {Landi},
  \citenamefont {Paternostro},\ and\ \citenamefont
  {Belenchia}}]{Landi-steady-state}%
  \BibitemOpen
  \bibfield  {author} {\bibinfo {author} {\bibfnamefont {G.~T.}\ \bibnamefont
  {Landi}}, \bibinfo {author} {\bibfnamefont {M.}~\bibnamefont {Paternostro}},\
  and\ \bibinfo {author} {\bibfnamefont {A.}~\bibnamefont {Belenchia}},\ }\href
  {http://arxiv.org/abs/2103.06247} {\  (\bibinfo {year} {2021})},\ \Eprint
  {https://arxiv.org/abs/2103.06247} {arXiv:2103.06247} \BibitemShut {NoStop}%
\bibitem [{\citenamefont {Sagawa}\ and\ \citenamefont
  {Ueda}(2009)}]{Sagawa2009b}%
  \BibitemOpen
  \bibfield  {author} {\bibinfo {author} {\bibfnamefont {T.}~\bibnamefont
  {Sagawa}}\ and\ \bibinfo {author} {\bibfnamefont {M.}~\bibnamefont {Ueda}},\
  }\href {https://doi.org/10.1103/PhysRevLett.102.250602} {\bibfield  {journal}
  {\bibinfo  {journal} {Phys. Rev. Lett.}\ }\textbf {\bibinfo {volume} {102}},\
  \bibinfo {pages} {250602} (\bibinfo {year} {2009})}\BibitemShut {NoStop}%
\bibitem [{\citenamefont {Jacobs}(2012)}]{Jacobs2012a}%
  \BibitemOpen
  \bibfield  {author} {\bibinfo {author} {\bibfnamefont {K.}~\bibnamefont
  {Jacobs}},\ }\href {https://doi.org/10.1103/PhysRevE.86.040106} {\bibfield
  {journal} {\bibinfo  {journal} {Phys. Rev. E}\ }\textbf {\bibinfo {volume}
  {86}},\ \bibinfo {pages} {040106(R)} (\bibinfo {year} {2012})}\BibitemShut
  {NoStop}%
\bibitem [{\citenamefont {Abdelkhalek}\ \emph {et~al.}(2016)\citenamefont
  {Abdelkhalek}, \citenamefont {Nakata},\ and\ \citenamefont
  {Reeb}}]{Abdelkhalek2016}%
  \BibitemOpen
  \bibfield  {author} {\bibinfo {author} {\bibfnamefont {K.}~\bibnamefont
  {Abdelkhalek}}, \bibinfo {author} {\bibfnamefont {Y.}~\bibnamefont
  {Nakata}},\ and\ \bibinfo {author} {\bibfnamefont {D.}~\bibnamefont {Reeb}},\
  }\href {http://arxiv.org/abs/1609.06981} {\  (\bibinfo {year} {2016})},\
  \Eprint {https://arxiv.org/abs/1609.06981} {arXiv:1609.06981} \BibitemShut
  {NoStop}%
\bibitem [{\citenamefont {Landauer}(1961)}]{Landauer1961}%
  \BibitemOpen
  \bibfield  {author} {\bibinfo {author} {\bibfnamefont {R.}~\bibnamefont
  {Landauer}},\ }\href {https://doi.org/10.1147/rd.53.0183} {\bibfield
  {journal} {\bibinfo  {journal} {IBM J. Res. Dev.}\ }\textbf {\bibinfo
  {volume} {5}},\ \bibinfo {pages} {261} (\bibinfo {year} {1961})}\BibitemShut
  {NoStop}%
\bibitem [{\citenamefont {Reeb}\ and\ \citenamefont {Wolf}(2014)}]{Reeb2013a}%
  \BibitemOpen
  \bibfield  {author} {\bibinfo {author} {\bibfnamefont {D.}~\bibnamefont
  {Reeb}}\ and\ \bibinfo {author} {\bibfnamefont {M.~M.}\ \bibnamefont
  {Wolf}},\ }\href {https://doi.org/10.1088/1367-2630/16/10/103011} {\bibfield
  {journal} {\bibinfo  {journal} {New J. Phys.}\ }\textbf {\bibinfo {volume}
  {16}},\ \bibinfo {pages} {103011} (\bibinfo {year} {2014})}\BibitemShut
  {NoStop}%
\bibitem [{\citenamefont {Mohammady}\ and\ \citenamefont
  {Romito}(2019)}]{Mohammady2019c}%
  \BibitemOpen
  \bibfield  {author} {\bibinfo {author} {\bibfnamefont {M.~H.}\ \bibnamefont
  {Mohammady}}\ and\ \bibinfo {author} {\bibfnamefont {A.}~\bibnamefont
  {Romito}},\ }\href {https://doi.org/10.22331/q-2019-08-19-175} {\bibfield
  {journal} {\bibinfo  {journal} {Quantum}\ }\textbf {\bibinfo {volume} {3}},\
  \bibinfo {pages} {175} (\bibinfo {year} {2019})}\BibitemShut {NoStop}%
\bibitem [{\citenamefont {Strasberg}(2019)}]{Strasberg2019}%
  \BibitemOpen
  \bibfield  {author} {\bibinfo {author} {\bibfnamefont {P.}~\bibnamefont
  {Strasberg}},\ }\href {https://doi.org/10.1103/PhysRevE.100.022127}
  {\bibfield  {journal} {\bibinfo  {journal} {Phys. Rev. E}\ }\textbf {\bibinfo
  {volume} {100}},\ \bibinfo {pages} {022127} (\bibinfo {year}
  {2019})}\BibitemShut {NoStop}%
\bibitem [{\citenamefont {Ozawa}(1984)}]{Ozawa1984}%
  \BibitemOpen
  \bibfield  {author} {\bibinfo {author} {\bibfnamefont {M.}~\bibnamefont
  {Ozawa}},\ }\href {https://doi.org/10.1063/1.526000} {\bibfield  {journal}
  {\bibinfo  {journal} {J. Math. Phys.}\ }\textbf {\bibinfo {volume} {25}},\
  \bibinfo {pages} {79} (\bibinfo {year} {1984})}\BibitemShut {NoStop}%
\bibitem [{\citenamefont {Elouard}\ \emph
  {et~al.}(2017{\natexlab{c}})\citenamefont {Elouard}, \citenamefont
  {Herrera-Mart{\'{i}}}, \citenamefont {Clusel},\ and\ \citenamefont
  {Auff{\`{e}}ves}}]{Elouard2017}%
  \BibitemOpen
  \bibfield  {author} {\bibinfo {author} {\bibfnamefont {C.}~\bibnamefont
  {Elouard}}, \bibinfo {author} {\bibfnamefont {D.~A.}\ \bibnamefont
  {Herrera-Mart{\'{i}}}}, \bibinfo {author} {\bibfnamefont {M.}~\bibnamefont
  {Clusel}},\ and\ \bibinfo {author} {\bibfnamefont {A.}~\bibnamefont
  {Auff{\`{e}}ves}},\ }\href {https://doi.org/10.1038/s41534-017-0008-4}
  {\bibfield  {journal} {\bibinfo  {journal} {npj Quantum Inf.}\ }\textbf
  {\bibinfo {volume} {3}},\ \bibinfo {pages} {9} (\bibinfo {year}
  {2017}{\natexlab{c}})}\BibitemShut {NoStop}%
\bibitem [{\citenamefont {Busch}\ and\ \citenamefont
  {Shimony}(1996)}]{Busch-Shimony-1997}%
  \BibitemOpen
  \bibfield  {author} {\bibinfo {author} {\bibfnamefont {P.}~\bibnamefont
  {Busch}}\ and\ \bibinfo {author} {\bibfnamefont {A.}~\bibnamefont
  {Shimony}},\ }\href {https://doi.org/10.1016/S1355-2198(96)00012-3}
  {\bibfield  {journal} {\bibinfo  {journal} {Stud. Hist. Philos. Sci. Part B
  Stud. Hist. Philos. Mod. Phys.}\ }\textbf {\bibinfo {volume} {27}},\ \bibinfo
  {pages} {397} (\bibinfo {year} {1996})}\BibitemShut {NoStop}%
\bibitem [{\citenamefont {Busch}(1998)}]{Busch1998a}%
  \BibitemOpen
  \bibfield  {author} {\bibinfo {author} {\bibfnamefont {P.}~\bibnamefont
  {Busch}},\ }\href {https://doi.org/10.1023/A:1026658532622} {\bibfield
  {journal} {\bibinfo  {journal} {Int. J. Theor. Phys.}\ }\textbf {\bibinfo
  {volume} {37}},\ \bibinfo {pages} {241} (\bibinfo {year} {1998})}\BibitemShut
  {NoStop}%
\bibitem [{\citenamefont {Yanase}(1961)}]{Yanase1961}%
  \BibitemOpen
  \bibfield  {author} {\bibinfo {author} {\bibfnamefont {M.~M.}\ \bibnamefont
  {Yanase}},\ }\href {https://doi.org/10.1103/PhysRev.123.666} {\bibfield
  {journal} {\bibinfo  {journal} {Phys. Rev.}\ }\textbf {\bibinfo {volume}
  {123}},\ \bibinfo {pages} {666} (\bibinfo {year} {1961})}\BibitemShut
  {NoStop}%
\bibitem [{\citenamefont {Ozawa}(2002)}]{Ozawa2002}%
  \BibitemOpen
  \bibfield  {author} {\bibinfo {author} {\bibfnamefont {M.}~\bibnamefont
  {Ozawa}},\ }\href {https://doi.org/10.1103/PhysRevLett.88.050402} {\bibfield
  {journal} {\bibinfo  {journal} {Phys. Rev. Lett.}\ }\textbf {\bibinfo
  {volume} {88}},\ \bibinfo {pages} {050402} (\bibinfo {year}
  {2002})}\BibitemShut {NoStop}%
\bibitem [{\citenamefont {Wigner}(1952)}]{E.Wigner1952}%
  \BibitemOpen
  \bibfield  {author} {\bibinfo {author} {\bibfnamefont {E.~P.}\ \bibnamefont
  {Wigner}},\ }\href {https://doi.org/10.1007/BF01948686} {\bibfield  {journal}
  {\bibinfo  {journal} {Zeitschrift f{\"{u}}r Phys. A Hadron. Nucl.}\ }\textbf
  {\bibinfo {volume} {133}},\ \bibinfo {pages} {101} (\bibinfo {year}
  {1952})}\BibitemShut {NoStop}%
\bibitem [{\citenamefont {Busch}()}]{Busch2010}%
  \BibitemOpen
  \bibfield  {author} {\bibinfo {author} {\bibfnamefont {P.}~\bibnamefont
  {Busch}},\ }\href {http://arxiv.org/abs/1012.4372} {\ }\Eprint
  {https://arxiv.org/abs/1012.4372} {arXiv:1012.4372} \BibitemShut {NoStop}%
\bibitem [{\citenamefont {Araki}\ and\ \citenamefont
  {Yanase}(1960)}]{Araki1960}%
  \BibitemOpen
  \bibfield  {author} {\bibinfo {author} {\bibfnamefont {H.}~\bibnamefont
  {Araki}}\ and\ \bibinfo {author} {\bibfnamefont {M.~M.}\ \bibnamefont
  {Yanase}},\ }\href {https://doi.org/10.1103/PhysRev.120.622} {\bibfield
  {journal} {\bibinfo  {journal} {Phys. Rev.}\ }\textbf {\bibinfo {volume}
  {120}},\ \bibinfo {pages} {622} (\bibinfo {year} {1960})}\BibitemShut
  {NoStop}%
\bibitem [{\citenamefont {Loveridge}\ \emph {et~al.}(2018)\citenamefont
  {Loveridge}, \citenamefont {Miyadera},\ and\ \citenamefont
  {Busch}}]{Loveridge2017a}%
  \BibitemOpen
  \bibfield  {author} {\bibinfo {author} {\bibfnamefont {L.}~\bibnamefont
  {Loveridge}}, \bibinfo {author} {\bibfnamefont {T.}~\bibnamefont
  {Miyadera}},\ and\ \bibinfo {author} {\bibfnamefont {P.}~\bibnamefont
  {Busch}},\ }\href {https://doi.org/10.1007/s10701-018-0138-3} {\bibfield
  {journal} {\bibinfo  {journal} {Found. Phys.}\ }\textbf {\bibinfo {volume}
  {48}},\ \bibinfo {pages} {135} (\bibinfo {year} {2018})}\BibitemShut
  {NoStop}%
\bibitem [{\citenamefont {Loveridge}(2020)}]{Loveridge2020a}%
  \BibitemOpen
  \bibfield  {author} {\bibinfo {author} {\bibfnamefont {L.}~\bibnamefont
  {Loveridge}},\ }\href {https://doi.org/10.1088/1742-6596/1638/1/012009}
  {\bibfield  {journal} {\bibinfo  {journal} {J. Phys. Conf. Ser.}\ }\textbf
  {\bibinfo {volume} {1638}},\ \bibinfo {pages} {012009} (\bibinfo {year}
  {2020})}\BibitemShut {NoStop}%
\bibitem [{\citenamefont {Wigner}\ and\ \citenamefont
  {Yanase}(1963)}]{Wigner1963}%
  \BibitemOpen
  \bibfield  {author} {\bibinfo {author} {\bibfnamefont {E.~P.}\ \bibnamefont
  {Wigner}}\ and\ \bibinfo {author} {\bibfnamefont {M.~M.}\ \bibnamefont
  {Yanase}},\ }\href {https://doi.org/10.1073/pnas.49.6.910} {\bibfield
  {journal} {\bibinfo  {journal} {Proc. Natl. Acad. Sci.}\ }\textbf {\bibinfo
  {volume} {49}},\ \bibinfo {pages} {910} (\bibinfo {year} {1963})}\BibitemShut
  {NoStop}%
\bibitem [{\citenamefont {Lieb}(1973)}]{Lieb1973}%
  \BibitemOpen
  \bibfield  {author} {\bibinfo {author} {\bibfnamefont {E.~H.}\ \bibnamefont
  {Lieb}},\ }\href {https://doi.org/10.1016/0001-8708(73)90011-X} {\bibfield
  {journal} {\bibinfo  {journal} {Adv. Math. (N. Y).}\ }\textbf {\bibinfo
  {volume} {11}},\ \bibinfo {pages} {267} (\bibinfo {year} {1973})}\BibitemShut
  {NoStop}%
\bibitem [{\citenamefont {Busch}\ \emph
  {et~al.}(1995{\natexlab{a}})\citenamefont {Busch}, \citenamefont
  {Grabowski},\ and\ \citenamefont {Lahti}}]{PaulBuschMarianGrabowski1995}%
  \BibitemOpen
  \bibfield  {author} {\bibinfo {author} {\bibfnamefont {P.}~\bibnamefont
  {Busch}}, \bibinfo {author} {\bibfnamefont {M.}~\bibnamefont {Grabowski}},\
  and\ \bibinfo {author} {\bibfnamefont {P.~J.}\ \bibnamefont {Lahti}},\ }\href
  {https://doi.org/10.1007/978-3-540-49239-9} {\emph {\bibinfo {title}
  {{Operational Quantum Physics}}}},\ \bibinfo {series} {Lecture Notes in
  Physics Monographs}, Vol.~\bibinfo {volume} {31}\ (\bibinfo  {publisher}
  {Springer Berlin Heidelberg},\ \bibinfo {address} {Berlin, Heidelberg},\
  \bibinfo {year} {1995})\BibitemShut {NoStop}%
\bibitem [{\citenamefont {Busch}\ \emph {et~al.}(1996)\citenamefont {Busch},
  \citenamefont {Lahti},\ and\ \citenamefont {{Peter
  Mittelstaedt}}}]{Busch1996}%
  \BibitemOpen
  \bibfield  {author} {\bibinfo {author} {\bibfnamefont {P.}~\bibnamefont
  {Busch}}, \bibinfo {author} {\bibfnamefont {P.~J.}\ \bibnamefont {Lahti}},\
  and\ \bibinfo {author} {\bibnamefont {{Peter Mittelstaedt}}},\ }\href
  {https://doi.org/10.1007/978-3-540-37205-9} {\emph {\bibinfo {title} {{The
  Quantum Theory of Measurement}}}},\ \bibinfo {series} {Lecture Notes in
  Physics Monographs}, Vol.~\bibinfo {volume} {2}\ (\bibinfo  {publisher}
  {Springer Berlin Heidelberg},\ \bibinfo {address} {Berlin, Heidelberg},\
  \bibinfo {year} {1996})\BibitemShut {NoStop}%
\bibitem [{\citenamefont {Mittelstaedt}(1997)}]{PeterMittelstaedt2004}%
  \BibitemOpen
  \bibfield  {author} {\bibinfo {author} {\bibfnamefont {P.}~\bibnamefont
  {Mittelstaedt}},\ }\href {https://doi.org/10.1017/CBO9780511564260} {\emph
  {\bibinfo {title} {{The Interpretation of Quantum Mechanics and the
  Measurement Process}}}}\ (\bibinfo  {publisher} {Cambridge University
  Press},\ \bibinfo {address} {Cambridge},\ \bibinfo {year} {1997})\BibitemShut
  {NoStop}%
\bibitem [{\citenamefont {Heinosaari}\ and\ \citenamefont
  {Ziman}(2011)}]{Heinosaari2011}%
  \BibitemOpen
  \bibfield  {author} {\bibinfo {author} {\bibfnamefont {T.}~\bibnamefont
  {Heinosaari}}\ and\ \bibinfo {author} {\bibfnamefont {M.}~\bibnamefont
  {Ziman}},\ }\href {https://doi.org/10.1017/CBO9781139031103} {\emph {\bibinfo
  {title} {{The Mathematical language of Quantum Theory}}}}\ (\bibinfo
  {publisher} {Cambridge University Press},\ \bibinfo {address} {Cambridge},\
  \bibinfo {year} {2011})\BibitemShut {NoStop}%
\bibitem [{\citenamefont {Holevo}(2011)}]{Holevo-Prob-Quantum}%
  \BibitemOpen
  \bibfield  {author} {\bibinfo {author} {\bibfnamefont {A.~S.}\ \bibnamefont
  {Holevo}},\ }\href@noop {} {\emph {\bibinfo {title} {{Probabilistic and
  Statistical Aspects of Quantum Theory}}}}\ (\bibinfo  {publisher} {Edizioni
  della Normale},\ \bibinfo {year} {2011})\BibitemShut {NoStop}%
\bibitem [{\citenamefont {Busch}\ \emph {et~al.}(2016)\citenamefont {Busch},
  \citenamefont {Lahti}, \citenamefont {Pellonp{\"{a}}{\"{a}}},\ and\
  \citenamefont {Ylinen}}]{Busch2016a}%
  \BibitemOpen
  \bibfield  {author} {\bibinfo {author} {\bibfnamefont {P.}~\bibnamefont
  {Busch}}, \bibinfo {author} {\bibfnamefont {P.}~\bibnamefont {Lahti}},
  \bibinfo {author} {\bibfnamefont {J.-P.}\ \bibnamefont
  {Pellonp{\"{a}}{\"{a}}}},\ and\ \bibinfo {author} {\bibfnamefont
  {K.}~\bibnamefont {Ylinen}},\ }\href
  {https://doi.org/10.1007/978-3-319-43389-9} {\emph {\bibinfo {title}
  {{Quantum Measurement}}}},\ Theoretical and Mathematical Physics\ (\bibinfo
  {publisher} {Springer International Publishing},\ \bibinfo {address} {Cham},\
  \bibinfo {year} {2016})\BibitemShut {NoStop}%
\bibitem [{\citenamefont {Hayashi}(2017)}]{Hayashi-QIT}%
  \BibitemOpen
  \bibfield  {author} {\bibinfo {author} {\bibfnamefont {M.}~\bibnamefont
  {Hayashi}},\ }\href {https://doi.org/10.1007/978-3-662-49725-8} {\emph
  {\bibinfo {title} {{Quantum Information Theory}}}},\ Graduate Texts in
  Physics\ (\bibinfo  {publisher} {Springer Berlin Heidelberg},\ \bibinfo
  {address} {Berlin, Heidelberg},\ \bibinfo {year} {2017})\BibitemShut
  {NoStop}%
\bibitem [{\citenamefont {Davies}\ and\ \citenamefont
  {Lewis}(1970)}]{Davies1970}%
  \BibitemOpen
  \bibfield  {author} {\bibinfo {author} {\bibfnamefont {E.~B.}\ \bibnamefont
  {Davies}}\ and\ \bibinfo {author} {\bibfnamefont {J.~T.}\ \bibnamefont
  {Lewis}},\ }\href {https://doi.org/10.1007/BF01647093} {\bibfield  {journal}
  {\bibinfo  {journal} {Commun. Math. Phys.}\ }\textbf {\bibinfo {volume}
  {17}},\ \bibinfo {pages} {239} (\bibinfo {year} {1970})}\BibitemShut
  {NoStop}%
\bibitem [{\citenamefont {Busch}\ \emph
  {et~al.}(1995{\natexlab{b}})\citenamefont {Busch}, \citenamefont
  {Grabowski},\ and\ \citenamefont {Lahti}}]{Busch1995}%
  \BibitemOpen
  \bibfield  {author} {\bibinfo {author} {\bibfnamefont {P.}~\bibnamefont
  {Busch}}, \bibinfo {author} {\bibfnamefont {M.}~\bibnamefont {Grabowski}},\
  and\ \bibinfo {author} {\bibfnamefont {P.~J.}\ \bibnamefont {Lahti}},\ }\href
  {https://doi.org/10.1007/BF02055331} {\bibfield  {journal} {\bibinfo
  {journal} {Found. Phys.}\ }\textbf {\bibinfo {volume} {25}},\ \bibinfo
  {pages} {1239} (\bibinfo {year} {1995}{\natexlab{b}})}\BibitemShut {NoStop}%
\bibitem [{\citenamefont {Busch}\ \emph {et~al.}(1990)\citenamefont {Busch},
  \citenamefont {Cassinelli},\ and\ \citenamefont {Lahti}}]{Busch1990}%
  \BibitemOpen
  \bibfield  {author} {\bibinfo {author} {\bibfnamefont {P.}~\bibnamefont
  {Busch}}, \bibinfo {author} {\bibfnamefont {G.}~\bibnamefont {Cassinelli}},\
  and\ \bibinfo {author} {\bibfnamefont {P.~J.}\ \bibnamefont {Lahti}},\ }\href
  {https://doi.org/10.1007/BF01889690} {\bibfield  {journal} {\bibinfo
  {journal} {Found. Phys.}\ }\textbf {\bibinfo {volume} {20}},\ \bibinfo
  {pages} {757} (\bibinfo {year} {1990})}\BibitemShut {NoStop}%
\bibitem [{\citenamefont {L{\"{u}}ders}(2006)}]{Luders2006}%
  \BibitemOpen
  \bibfield  {author} {\bibinfo {author} {\bibfnamefont {G.}~\bibnamefont
  {L{\"{u}}ders}},\ }\href {https://doi.org/10.1002/andp.200610207} {\bibfield
  {journal} {\bibinfo  {journal} {Ann. Phys.}\ }\textbf {\bibinfo {volume}
  {15}},\ \bibinfo {pages} {663} (\bibinfo {year} {2006})}\BibitemShut
  {NoStop}%
\bibitem [{\citenamefont {Lahti}\ \emph {et~al.}(1991)\citenamefont {Lahti},
  \citenamefont {Busch},\ and\ \citenamefont {Mittelstaedt}}]{Lahti1991}%
  \BibitemOpen
  \bibfield  {author} {\bibinfo {author} {\bibfnamefont {P.~J.}\ \bibnamefont
  {Lahti}}, \bibinfo {author} {\bibfnamefont {P.}~\bibnamefont {Busch}},\ and\
  \bibinfo {author} {\bibfnamefont {P.}~\bibnamefont {Mittelstaedt}},\ }\href
  {https://doi.org/10.1063/1.529504} {\bibfield  {journal} {\bibinfo  {journal}
  {J. Math. Phys.}\ }\textbf {\bibinfo {volume} {32}},\ \bibinfo {pages} {2770}
  (\bibinfo {year} {1991})}\BibitemShut {NoStop}%
\bibitem [{\citenamefont {Zurek}(2003{\natexlab{b}})}]{Zurek2003}%
  \BibitemOpen
  \bibfield  {author} {\bibinfo {author} {\bibfnamefont {W.~H.}\ \bibnamefont
  {Zurek}},\ }\href {https://doi.org/10.1103/RevModPhys.75.715} {\bibfield
  {journal} {\bibinfo  {journal} {Rev. Mod. Phys.}\ }\textbf {\bibinfo {volume}
  {75}},\ \bibinfo {pages} {715} (\bibinfo {year}
  {2003}{\natexlab{b}})}\BibitemShut {NoStop}%
\bibitem [{\citenamefont {Ghirardi}\ \emph {et~al.}(1986)\citenamefont
  {Ghirardi}, \citenamefont {Rimini},\ and\ \citenamefont
  {Weber}}]{Ghirardi1986}%
  \BibitemOpen
  \bibfield  {author} {\bibinfo {author} {\bibfnamefont {G.~C.}\ \bibnamefont
  {Ghirardi}}, \bibinfo {author} {\bibfnamefont {A.}~\bibnamefont {Rimini}},\
  and\ \bibinfo {author} {\bibfnamefont {T.}~\bibnamefont {Weber}},\ }\href
  {https://doi.org/10.1103/PhysRevD.34.470} {\bibfield  {journal} {\bibinfo
  {journal} {Phys. Rev. D}\ }\textbf {\bibinfo {volume} {34}},\ \bibinfo
  {pages} {470} (\bibinfo {year} {1986})}\BibitemShut {NoStop}%
\bibitem [{\citenamefont {Allahverdyan}\ and\ \citenamefont
  {Nieuwenhuizen}(2005)}]{Allahverdyan2005}%
  \BibitemOpen
  \bibfield  {author} {\bibinfo {author} {\bibfnamefont {A.~E.}\ \bibnamefont
  {Allahverdyan}}\ and\ \bibinfo {author} {\bibfnamefont {T.~M.}\ \bibnamefont
  {Nieuwenhuizen}},\ }\href {https://doi.org/10.1103/PhysRevE.71.066102}
  {\bibfield  {journal} {\bibinfo  {journal} {Phys. Rev. E}\ }\textbf {\bibinfo
  {volume} {71}},\ \bibinfo {pages} {066102} (\bibinfo {year}
  {2005})}\BibitemShut {NoStop}%
\bibitem [{\citenamefont {Allahverdyan}(2014)}]{Allahverdyan2014}%
  \BibitemOpen
  \bibfield  {author} {\bibinfo {author} {\bibfnamefont {A.~E.}\ \bibnamefont
  {Allahverdyan}},\ }\href {https://doi.org/10.1103/PhysRevE.90.032137}
  {\bibfield  {journal} {\bibinfo  {journal} {Phys. Rev. E}\ }\textbf {\bibinfo
  {volume} {90}},\ \bibinfo {pages} {032137} (\bibinfo {year}
  {2014})}\BibitemShut {NoStop}%
\bibitem [{\citenamefont {Esposito}\ \emph {et~al.}(2009)\citenamefont
  {Esposito}, \citenamefont {Harbola},\ and\ \citenamefont
  {Mukamel}}]{Esposito2009}%
  \BibitemOpen
  \bibfield  {author} {\bibinfo {author} {\bibfnamefont {M.}~\bibnamefont
  {Esposito}}, \bibinfo {author} {\bibfnamefont {U.}~\bibnamefont {Harbola}},\
  and\ \bibinfo {author} {\bibfnamefont {S.}~\bibnamefont {Mukamel}},\ }\href
  {https://doi.org/10.1103/RevModPhys.81.1665} {\bibfield  {journal} {\bibinfo
  {journal} {Rev. Mod. Phys.}\ }\textbf {\bibinfo {volume} {81}},\ \bibinfo
  {pages} {1665} (\bibinfo {year} {2009})}\BibitemShut {NoStop}%
\bibitem [{\citenamefont {Campisi}\ \emph {et~al.}(2011)\citenamefont
  {Campisi}, \citenamefont {H{\"{a}}nggi},\ and\ \citenamefont
  {Talkner}}]{Campisi2011}%
  \BibitemOpen
  \bibfield  {author} {\bibinfo {author} {\bibfnamefont {M.}~\bibnamefont
  {Campisi}}, \bibinfo {author} {\bibfnamefont {P.}~\bibnamefont
  {H{\"{a}}nggi}},\ and\ \bibinfo {author} {\bibfnamefont {P.}~\bibnamefont
  {Talkner}},\ }\href {https://doi.org/10.1103/RevModPhys.83.771} {\bibfield
  {journal} {\bibinfo  {journal} {Rev. Mod. Phys.}\ }\textbf {\bibinfo {volume}
  {83}},\ \bibinfo {pages} {771} (\bibinfo {year} {2011})}\BibitemShut
  {NoStop}%
\bibitem [{\citenamefont {Deffner}\ \emph {et~al.}(2016)\citenamefont
  {Deffner}, \citenamefont {Paz},\ and\ \citenamefont {Zurek}}]{Deffner2016a}%
  \BibitemOpen
  \bibfield  {author} {\bibinfo {author} {\bibfnamefont {S.}~\bibnamefont
  {Deffner}}, \bibinfo {author} {\bibfnamefont {J.~P.}\ \bibnamefont {Paz}},\
  and\ \bibinfo {author} {\bibfnamefont {W.~H.}\ \bibnamefont {Zurek}},\ }\href
  {https://doi.org/10.1103/PhysRevE.94.010103} {\bibfield  {journal} {\bibinfo
  {journal} {Phys. Rev. E}\ }\textbf {\bibinfo {volume} {94}},\ \bibinfo
  {pages} {010103(R)} (\bibinfo {year} {2016})}\BibitemShut {NoStop}%
\bibitem [{\citenamefont {Luo}(2005)}]{Luo2005}%
  \BibitemOpen
  \bibfield  {author} {\bibinfo {author} {\bibfnamefont {S.~L.}\ \bibnamefont
  {Luo}},\ }\href {https://doi.org/10.1007/s11232-005-0098-6} {\bibfield
  {journal} {\bibinfo  {journal} {Theor. Math. Phys.}\ }\textbf {\bibinfo
  {volume} {143}},\ \bibinfo {pages} {681} (\bibinfo {year}
  {2005})}\BibitemShut {NoStop}%
\bibitem [{\citenamefont {Holevo}(1973)}]{Holevo1973}%
  \BibitemOpen
  \bibfield  {author} {\bibinfo {author} {\bibfnamefont {A.~S.}\ \bibnamefont
  {Holevo}},\ }\href@noop {} {\bibfield  {journal} {\bibinfo  {journal} {Probl.
  Inf. Transm.}\ }\textbf {\bibinfo {volume} {9}},\ \bibinfo {pages} {177}
  (\bibinfo {year} {1973})}\BibitemShut {NoStop}%
\bibitem [{\citenamefont {Audenaert}(2014)}]{Audenaert2014}%
  \BibitemOpen
  \bibfield  {author} {\bibinfo {author} {\bibfnamefont {K.~M.~R.}\
  \bibnamefont {Audenaert}},\ }\href {https://doi.org/10.1063/1.4901039}
  {\bibfield  {journal} {\bibinfo  {journal} {J. Math. Phys.}\ }\textbf
  {\bibinfo {volume} {55}},\ \bibinfo {pages} {112202} (\bibinfo {year}
  {2014})}\BibitemShut {NoStop}%
\bibitem [{\citenamefont {Marvian}\ and\ \citenamefont
  {Spekkens}(2014)}]{Marvian2014}%
  \BibitemOpen
  \bibfield  {author} {\bibinfo {author} {\bibfnamefont {I.}~\bibnamefont
  {Marvian}}\ and\ \bibinfo {author} {\bibfnamefont {R.~W.}\ \bibnamefont
  {Spekkens}},\ }\href {https://doi.org/10.1038/ncomms4821} {\bibfield
  {journal} {\bibinfo  {journal} {Nat. Commun.}\ }\textbf {\bibinfo {volume}
  {5}},\ \bibinfo {pages} {3821} (\bibinfo {year} {2014})}\BibitemShut
  {NoStop}%
\bibitem [{\citenamefont {Takagi}(2019)}]{Takagi2018}%
  \BibitemOpen
  \bibfield  {author} {\bibinfo {author} {\bibfnamefont {R.}~\bibnamefont
  {Takagi}},\ }\href {https://doi.org/10.1038/s41598-019-50279-w} {\bibfield
  {journal} {\bibinfo  {journal} {Sci. Rep.}\ }\textbf {\bibinfo {volume}
  {9}},\ \bibinfo {pages} {14562} (\bibinfo {year} {2019})}\BibitemShut
  {NoStop}%
\bibitem [{\citenamefont {Ozawa}(2001)}]{Ozawa2001}%
  \BibitemOpen
  \bibfield  {author} {\bibinfo {author} {\bibfnamefont {M.}~\bibnamefont
  {Ozawa}},\ }\href {https://doi.org/10.1103/PhysRevA.63.032109} {\bibfield
  {journal} {\bibinfo  {journal} {Phys. Rev. A}\ }\textbf {\bibinfo {volume}
  {63}},\ \bibinfo {pages} {032109} (\bibinfo {year} {2001})}\BibitemShut
  {NoStop}%
\bibitem [{\citenamefont {Haapasalo}\ \emph {et~al.}(2011)\citenamefont
  {Haapasalo}, \citenamefont {Lahti},\ and\ \citenamefont
  {Schultz}}]{Haapasalo2011}%
  \BibitemOpen
  \bibfield  {author} {\bibinfo {author} {\bibfnamefont {E.}~\bibnamefont
  {Haapasalo}}, \bibinfo {author} {\bibfnamefont {P.}~\bibnamefont {Lahti}},\
  and\ \bibinfo {author} {\bibfnamefont {J.}~\bibnamefont {Schultz}},\ }\href
  {https://doi.org/10.1103/PhysRevA.84.052107} {\bibfield  {journal} {\bibinfo
  {journal} {Phys. Rev. A}\ }\textbf {\bibinfo {volume} {84}},\ \bibinfo
  {pages} {052107} (\bibinfo {year} {2011})}\BibitemShut {NoStop}%
\end{thebibliography}%


\widetext
\appendix

\section{Sufficient conditions for a vanishing average  heat}\label{app:zero-average-heat}
We may rewrite the average heat, shown in \eq{eq:average-heat-energy}, as 
\begin{align*}
\avg{\qq} = \tr[H\sub{\aa} (\jj_\xx(\eta) - \eta)] = \tr[(\jj^*_\xx(H\sub{\aa}) - H\sub{\aa}) \eta]. 
\end{align*}
Here $\eta:= \tr\sub{\s}[U(\rho \otimes \xi)U^\dagger]$ is the reduced state of the apparatus  after premeasurement, and $\jj_\xx$ is the channel induced by the $\Z$-instrument $\jj$. In order to guarantee that $\avg{\qq}=0$ for all  $\eta$ (and hence all $\rho$), then $H\sub{\aa}$ must be a fixed point of the dual channel $\jj^*_\xx$, i.e., we must have $\jj^*_\xx(H\sub{\aa}) = H\sub{\aa}$. Throughout what follows, we shall assume that $\Z$ is a sharp observable that satisfies the Yanase condition, i.e.,  commutes with $H\sub{\aa}$.  Now assume that $\Z$  is implemented by the L\"uders instrument  $\jj^L_x(\cdot) \equiv {\jj^L_x}^*(\cdot)  = \Z_x  (\cdot) \Z_x $ (which is repeatable if and only if $\Z$ is sharp). It trivially follows that $\avg{\qq} = 0$, since ${\jj^L_\xx}^*(H\sub{\aa}) = \sum_x \Z_x  H\sub{\aa} \Z_x  = \sum_x \Z_x  H\sub{\aa} =  H\sub{\aa}$. If $\Z$ is instead implemented by an arbitrary  repeatable $\Z$-instrument $\jj$, then a sufficient condition for a vanishing average heat is if $H\sub{\aa} = \sum_x \epsilon_x \Z_x$. To see this, note from \eq{eq:repeatable-defn} that repeatability of $\jj$ implies that $\jj_y^*(\Z_x) = \delta_{x,y} \Z_x$. But this implies that $\jj_\xx^*(\Z_x) = \Z_x$. Therefore, we have
\begin{align*}
\jj^*_\xx(H\sub{\aa}) &= \sum_{x \in \xx} \epsilon_x \jj^*_\xx(\Z_x) =   \sum_{x\in \xx}\epsilon_x \Z_x  = H\sub{\aa}.
\end{align*}

To see that repeatability and the Yanase condition alone do not guarantee that $\avg{\qq} = 0$, let us consider the following simple example where $\ha \simeq \co^{2N}$ is finite-dimensional, where we identify the value space of $\Z$ as $\xx = \{1,\dots, N\}$, and where the Hamiltonian can be written in a diagonal form $H\sub{\aa} = \sum_i \epsilon_i P_i$, with $\{P_i\}$ an ortho-complete set of rank-1  projection operators. Assume that for each $x\in \xx$, we have $\Z_x = P_{2x - 1} + P_{2x}$.  Clearly, $\Z_x$ commutes with $H\sub{\aa}$.  Now note that for discrete sharp observables $\Z$, all $\Z$-instruments $\jj$ can be constructed as a sequential operation 
\begin{align}\label{eq:sequential-sharp-instrument}
&\jj_x(T) = \Phi(\Z_x T \Z_x), &\jj_x^*(B) = \Z_x\Phi^*(B) \Z_x,
\end{align} 
to hold for all $x\in \xx$, $T \in \trc(\ha)$, and $B \in \lo(\ha)$,  where $\Phi : \trc(\ha) \to \trc(\ha)$ is an arbitrary channel  \cite{Ozawa2001}. Now let us also assume that  $\Phi^*$ acts as a   ``depolarising channel'' on each  eigenvalue-1 eigenspace of $\Z_x $, that is, 
\begin{align}\label{eq:sharp-instrument-depolarising-channel}
\Phi^*(B) = \sum_{x\in \xx} \frac{\tr[B \Z_x]}{2} \Z_x
\end{align}
 for all $B \in \lo(\ha)$. To verify that \eq{eq:sharp-instrument-depolarising-channel} satisfies the repeatability condition, we use \eq{eq:sequential-sharp-instrument} to compute 
\begin{align*}
\jj_x^*(\Z_y) =  \Z_x\Phi^*(\Z_y) \Z_x = \frac{\tr[\Z_y \Z_x]}{2} \Z_x  = \delta_{x,y} \Z_x. 
\end{align*}
But, we now have 
\begin{align*}
\jj^*_\xx(H\sub{\aa}) =\sum_{x\in \xx} \Z_x \Phi^*(H\sub{\aa})\Z_x =  \sum_{x\in \xx} \frac{\tr[H\sub{\aa} \Z_x]}{2} \Z_x = \sum_{x\in \xx} \frac{1}{2}(\epsilon_{2x-1} + \epsilon_{2x})\Z_x,
\end{align*}
 which equals   $H\sub{\aa}$ only if $\epsilon_{2x-1} = \epsilon_{2x}$ for all $x\in \xx$.

\section{Variance of  heat}\label{app:heat-variance}

As shown in \eq{eq:heat}, the   measurement scheme $\mm:= (\ha, \xi, U, \Z)$ for the $\E$-instrument $\ii$ produces the heat
\begin{align*}
\qq(x) := \Delta \e(x)  - \ww =  \tr[H (\sigma_x  - \varrho')]
\end{align*}
for all input states $\rho\in \s(\hs)$ and outcomes $x\in \xx$ which occur with probability $p^\E_\rho (x)   := \tr[\E_x  \rho] >0$. Here, we define $\varrho' := U(\rho \otimes \xi)U^\dagger$ as the joint premeasured state of system and apparatus, and $\sigma_x$ as the normalised conditional  state of the composite system defined in  \eq{eq:joint-state-obj}. The average heat is thus  
\begin{align*}
\avg{\qq} := \sum_{x\in \xx} p^\E_\rho (x)   \qq(x) = \tr[H (\overline{\bm{\sigma}} - \varrho')],
\end{align*}
where $\overline{\bm{\sigma}}:= \sum_{x\in \xx} p^\E_\rho (x)   \sigma_x$, as defined in  \eq{eq:joint-state-obj-average}. The variance of heat is $\var{\qq} := \avg{\qq^2} - \avg{\qq}^2$, where $\avg{\qq^2} := \sum_{x\in \xx} p^\E_\rho (x)   \qq(x)^2$. Each term reads 
\begin{align*}
\avg{\qq}^2 &= \tr[H \overline{\bm{\sigma}}]^2 -2 \tr[H \overline{\bm{\sigma}}] \tr[H  \varrho'] + \tr[H  \varrho']^2,\\
\avg{\qq^2} &=  \sum_{x\in \xx} p^\E_\rho (x)  \bigg( \tr[H \sigma_x ]^2 -2\tr[H \sigma_x ] \tr[ H \varrho'] + \tr[ H \varrho']^2 \bigg) , \nonumber \\
&=  \bigg(\sum_{x\in \xx} p^\E_\rho (x)  \tr[H \sigma_x ]^2 \bigg)- 2\tr[H \overline{\bm{\sigma}}] \tr[ H \varrho'] + \tr[ H \varrho']^2.
\end{align*}
Here, in the final line we have used the fact that $\sum_x p^\E_\rho (x) \tr[H \sigma_x ] \tr[ H \varrho'] = \tr[H \overline{\bm{\sigma}}] \tr[ H \varrho']$. Therefore, the variance of heat will  be 
\begin{align*}
\var{\qq} := \avg{\qq^2} - \avg{\qq}^2 &= \bigg(\sum_{x\in \xx} p^\E_\rho (x)  \tr[H \sigma_x ]^2 \bigg) - \tr[H \overline{\bm{\sigma}}]^2, \nonumber \\
 &= \bigg(\sum_{x\in \xx} p^\E_\rho (x)  \tr[H \sigma_x ]^2 \bigg) - \tr[H \overline{\bm{\sigma}}]^2 + \tr[H^2 \overline{\bm{\sigma}}] - \tr[H^2 \overline{\bm{\sigma}}], \nonumber \\
& = \mathrm{V}(H,  \overline{\bm{\sigma}}) - \sum_{x\in \xx} p^\E_\rho (x)   \mathrm{V}(H,  \sigma_x ),
\end{align*}
where again we note that $\sum_x p^\E_\rho (x) \tr[H^2 \sigma_x ]  = \tr[H^2 \overline{\bm{\sigma}}]$, and  we recall that for any self-adjoint $A\in \lo(\h)$ and $\varrho \in \s(\h)$, $\mathrm{V}(A, \varrho) := \tr[A^2 \varrho] - \tr[A \varrho]^2$ is the variance of $A$ in $\varrho$.

\section{Skew information identity}\label{app:skew-info-identity}
Let us consider a compound system $\h\sub{1} \otimes \h\sub{2}$, with the additive Hamiltonian $H = H\sub{1} \otimes \one\sub{2} + \one\sub{1}\otimes H\sub{2}$, prepared in the state 
\begin{align*}
\varrho := \sum_i p_i P_i \otimes \sigma_i,
\end{align*}
where $p_i$ is a probability distribution,  $\{\ket{i}\}$ is an orthonormal basis of $\h\sub{1}$ with $P_i \equiv |i\>\<i|$, and $\sigma_i \in \s(\h\sub{2})$ are arbitrary states. We shall denote $\tau := \sum_i p_i P_i \equiv \tr\sub{2}[\varrho]$, and $ \overline{\bm{\sigma}} := \sum_i p_i \sigma_i \equiv \tr\sub{1}[\varrho]$. The Wigner-Yanase-Dyson skew information of $\varrho$, with reference to $H$, is $\qv(H, \varrho) := \tr[H ^2 \varrho] - \tr[H \varrho^\alpha H \varrho^{1-\alpha} H]$, with $\alpha \in (0,1)$. The first term reads
\begin{align}\label{eq:skew-info-bipartite-1}
\tr[H ^2 \varrho] =& \sum_i p_i \tr[(H\sub{1}^2 \otimes \one\sub{2} + \one\sub{1}\otimes H\sub{2}^2 + 2 H\sub{1} \otimes H\sub{2}) P_i \otimes \sigma_x ], \nonumber \\
& = \tr[H\sub{1}^2 \tau] + \tr[H\sub{2} ^2 \overline{\bm{\sigma}}] + 2 \sum_i p_i \tr[H\sub{1} P_i] \tr[H\sub{2} \sigma_i]. 
\end{align}
Here, we use the definition of the partial trace, together with the fact that $\tr[T_1  \otimes T_2] = \tr[T_1] \tr[T_2]$ for all $T_1 \in \trc(\h\sub{1})$ and $T_2 \in \trc(\h\sub{2})$.

Now let us assume that $\ket{i}$ are eigenstates of $H\sub{1}$, so that $[P_i, H\sub{1}]=\zero$. Given that $\{P_i \otimes \sigma_i\}$ are pairwise orthogonal, we have $\varrho^\alpha = \sum_i p_i^\alpha P_i \otimes \sigma_i^\alpha$ for all $\alpha \in (0,1)$. Therefore, we may write 
\begin{align}\label{eq:skew-info-bipartite-2}
\tr[H \varrho^\alpha H \varrho^{1-\alpha} ] & = \sum_{i,j} p_i^\alpha p_j^{1-\alpha} \tr[(H\sub{1} \otimes \one\sub{2} + \one\sub{1}\otimes H\sub{2}) P_i \otimes \sigma_i^\alpha(H\sub{1} \otimes \one\sub{2} + \one\sub{1}\otimes H\sub{2})P_j \otimes \sigma_j^{1-\alpha}], \nonumber \\
& = \sum_i p_i \tr[(H\sub{1} \otimes \one\sub{2}) P_i \otimes \sigma_i^\alpha (H\sub{1} \otimes \one\sub{2}) P_i \otimes \sigma_i^{1-\alpha}] \nonumber  \\
& \qquad  + \sum_i p_i \tr[(\one\sub{1}\otimes H\sub{2})P_i \otimes \sigma_i^\alpha (\one\sub{1}\otimes H\sub{2})P_i \otimes \sigma_i^{1-\alpha}] \nonumber  \\
& \qquad + \sum_i p_i \tr[(H\sub{1} \otimes \one\sub{2})P_i \otimes \sigma_i^\alpha (\one\sub{1}\otimes H\sub{2})P_i \otimes \sigma_i^{1-\alpha}] \nonumber \\
& \qquad  + \sum_i p_i \tr[(\one\sub{1}\otimes H\sub{2})P_i \otimes \sigma_i^\alpha (H\sub{1} \otimes \one\sub{2})P_i \otimes \sigma_i^{1-\alpha}], \nonumber \\
& = \sum_i p_i \tr[H\sub{1}^2 P_i \otimes \sigma_i] + \sum_i p_i \tr[P_i \otimes H\sub{2} \sigma_i^\alpha H\sub{2} \sigma_i^{1-\alpha}] + 2\sum_i p_i \tr[H\sub{1} P_i \otimes \sigma_i^\alpha H\sub{2} \sigma_i^{1-\alpha}] , \nonumber \\
& = \tr[H\sub{1}^2 \tau] + \sum_i p_i \tr[H\sub{2} \sigma_i^\alpha H\sub{2} \sigma_i^{1-\alpha}] + 2 \sum_i p_i \tr[H\sub{1} P_i] \tr[H\sub{2} \sigma_i].
\end{align}
Here, in the second line we have used the fact that $P_i$ commute with $H\sub{1}$ which implies that the trace vanishes if $i \ne j$. By combining \eq{eq:skew-info-bipartite-1} and \eq{eq:skew-info-bipartite-2}, we thus observe that 
\begin{align*}
\qv(H, \varrho) &= \tr[H\sub{2} ^2 \overline{\bm{\sigma}}]  - \sum_i p_i \tr[H\sub{2} \sigma_i^\alpha H\sub{2} \sigma_i^{1-\alpha}], \nonumber \\
& = \sum_i p_i \bigg( \tr[H\sub{2} ^2 \sigma_i ]  - \tr[H\sub{2} \sigma_i^\alpha H\sub{2} \sigma_i^{1-\alpha}]\bigg) = \sum_i p_i \qv(H\sub{2}, \sigma_i), 
\end{align*}
where in the final line we use $\tr[H\sub{2} ^2 \overline{\bm{\sigma}}] = \sum_i p_i \tr[H\sub{2} ^2 \sigma_i ]$.

\section{Relation to ``conditional'' change in energy}\label{app:conditional-energy}

In the main text, we remained silent as to the interpretation of the quantum state $\rho \in \s(\hs)$, in terms of which work and heat have been defined. Let us assume that such a  state is to be understood as a classical ensemble $\{p_k, \rho_k\}$, where $\rho_k$ are different state preparations that are sampled by the probability distribution $p_k$ such that $\rho = \sum_k p_k \rho_k$. The premeasurement work $\ww$, defined in \eq{eq:work}, can thus be understood as the average $\ww = \sum_k p_k \ww_k$, with  $\ww_k$ the work for the preparation $\rho_k$.  However, such an interpretation is not afforded for the change in internal energy as defined in  \eq{eq:increase-energy} and \eq{eq:increase-energy-reduced}. That is, $\Delta \e(x)$, $\Delta \e\sub{\s}(x)$, and $\Delta \e\sub{\aa}(x)$ are not, in general, the average change in internal energy  for the different state preparations. The reason for this is that while the final energies are conditional on the measurement outcome $x$, the initial energies are not. 

To illustrate this, let   us only consider the energy change of the measured system $\s$.  We first define  $p(x|k) \equiv p^E_{\rho_k}(x) := \tr[\E_x  \rho_k]$ as the conditional probability of observing outcome $x$ given state preparation $\rho_k$. Given that the total probability of observing outcome $x$, for all preparations $\rho_k$,  is $p(x) := \sum_k  p_k p(x|k) = \sum_k p_k \tr[\E_x  \rho_k] = \tr[\E_x  \rho] \equiv p^\E_\rho (x)$, we may use Bayes' theorem to obtain the probability of state preparation $k$ given observation of  outcome $x$ as $p(k|x) =  p_k  p(x|k) / p(x) \equiv    p_k \tr[\E_x  \rho_k] / \tr[\E_x  \rho]  $. Now let us define by $\rho_{(x|k)} := \ii_x(\rho_k) / p(x|k)$  the conditional post-measurement state of the system, given the initial preparation  $\rho_k$, such that $p(x|k) >0$. The change in internal energy, given outcome $x$ for preparation $k$, will thus be  $\Delta \e\sub{\s}(x|k):= \tr[H\sub{\s}(\rho_{(x|k)} - \rho_k)]$.    On the other hand, the  ``weighted average'' of $\Delta \e\sub{\s}(x |k)$, over the initial preparations $\rho_k$, will be 
\begin{align}
\sum_k p(k|x) \Delta \e\sub{\s}(x|k)  &= \sum_k  \frac{ p_k \tr[\E_x  \rho_k]  }{\tr[\E_x  \rho]} \left( \frac{\tr[H\sub{\s} \ii_x(\rho_k)]}{\tr[\E_x  \rho_k]} - \tr[H\sub{\s} \rho_k] \right), \nonumber \\
&  =  \frac{   \tr[H\sub{\s} \ii_x(\rho)] }{\tr[\E_x  \rho]  } - \sum_k \frac{ p_k \tr[\E_x  \rho_k]  }{\tr[\E_x  \rho]}\tr[H\sub{\s} \rho_k], \nonumber \\
&  = \tr\left[H\sub{\s} \left(\rho_x   - \frac{\sum_k\tr[\E_x  \rho_k] p_k\rho_k}{\tr[\E_x  \rho]  }\right)\right].
\end{align}
But this does not recover $\Delta \e\sub{\s}(x) = \tr[H\sub{\s}(\rho_x   - \rho)]$ in general, except  when for all $k$,  either: (i) $\rho_k = \rho$; (ii) $\tr[\E_x \rho_k]= \tr[\E_x \rho]$; or (iii) $\tr[H\sub{\s} \rho_k] = \tr[H\sub{\s} \rho]$.

Consequently, we are left with three options. We must either: (a)  restrict our analysis only to ensembles that satisfy one of (i)-(iii) above; (b) use an alternative to Bayesian probability theory to take weighted averages of $\Delta \e\sub{\s}(x|k)$; or (c)  modify the definition of $\Delta \e\sub{\s}$ so that it is fully conditional on the measurement outcome $x$. In the present manuscript we choose option (a)-(i), namely, we do not consider $\rho$ as an ensemble, but rather as an irreducible state defined by a given preparation procedure. However, in Ref. \cite{Mohammady2019c} we took option (c) and defined the fully conditional change in internal energy of the system, given that outcome $x$ is observed with non-zero probability,  as
\begin{align}\label{eq:conditional-energy-change-defn}
\Delta \tilde\e\sub{\s}(x) &:= \frac{\tr[H\sub{\s}\ii_x(\rho)]}{p^\E_\rho (x)  } - \frac{\frac{1}{2}\tr[\ii_x(H\sub{\s} \rho + \rho H\sub{\s})]}{p^\E_\rho (x)  } \equiv \frac{\tr[H\sub{\s}\ii_x(\rho)]}{p^\E_\rho (x)  } - \frac{\frac{1}{2}\tr[\E_x (H\sub{\s} \rho + \rho H\sub{\s})]}{p^\E_\rho (x)  }, \nonumber \\
&= \tr[H\sub{\s}\rho_x] - \frac{\frac{1}{2}\tr[\E_x (H\sub{\s} \rho + \rho H\sub{\s})]}{p^\E_\rho (x)  }.
\end{align}
As before, for any $x$ and $\rho$ such that $p^\E_\rho (x) =0$, we define $\Delta \tilde \e\sub{\s}(x):=0$. The equality on the first line follows from the fact that for any $T \in \trc(\hs)$, $\tr[\ii_x(T)] = \tr[\ii^*_x(\onesys) T] = \tr[\E_x T]$. Here, the second term is the real component of the generalised weak value of $H\sub{\s}$, given state $\rho$, and post-selected by outcome $x$ of the observable $\E$ \cite{Haapasalo2011}; the second term is the initial energy of the system,  \emph{conditioned}  on subsequently observing outcome $x$.   It can easily be verified that $\sum_k p(k|x) \Delta \tilde \e\sub{\s}(x|k)  = \Delta \tilde \e\sub{\s}(x)$ for all possible ensembles $\{p_k, \rho_k\}$ such that $\sum_k p_k \rho_k = \rho$. 

For a measurement scheme $\mm:= (\ha, \xi, U, \Z)$, we may use the definition in \eq{eq:conditional-energy-change-defn} to write the total conditional change in energy of the composite system  $\h := \hs\otimes \ha$,  with additive Hamiltonian $H := H\sub{\s} \otimes \oneapp + \onesys \otimes H\sub{\aa}$,  as 
\begin{align}\label{eq:conditional-energy-change-tot}
\Delta \tilde \e(x) &:= \frac{\tr[H \Phi_x(\rho \otimes  \xi )]}{p^\E_\rho (x)  }  - \frac{\frac{1}{2}\tr[\Phi_x(H\rho \otimes  \xi  + \rho \otimes  \xi H )]}{p^\E_\rho (x)  }\equiv  \frac{\tr[H \Phi_x(\rho \otimes  \xi )]}{p^\E_\rho (x)  }  - \frac{\frac{1}{2}\tr[\mathcal{Z}_x (H\rho \otimes  \xi  + \rho \otimes  \xi H )]}{p^\E_\rho (x)  }, \nonumber \\
& \equiv \tr[H \sigma_x] - \frac{\frac{1}{2}\tr[\mathcal{Z}_x (H\rho \otimes  \xi  + \rho \otimes  \xi H )]}{p^\E_\rho (x)  }.
\end{align}
Here, we define the operation $\Phi_x : \trc(\h) \to \trc(\h), T \mapsto  \idchsys\otimes \jj_x(U T U^\dagger)$, where $\idchsys$ denotes the identity channel acting on $\hs$, and $\jj$ is a $\Z$-instrument on $\ha$. Therefore, $\sigma_x = \Phi_x(\rho \otimes  \xi)/p^\E_\rho (x)$. We also define $\mathcal{Z}_x := U^\dagger(\onesys \otimes \Z_x ) U \equiv \Phi_x^*(\one)$ as the Heisenberg evolved pointer observable. 

As discussed in the main text, pointer objectification requires that $\Z$ be implemented by a repeatable instrument $\jj$, and stability of the objectified values demands that the pointer observable satisfies  the Yanase condition $[\Z, H] = [\Z, H\sub{\aa}] = \zero$. We shall make these assumptions, together with sharpness of $\Z$,  throughout what follows. 

Let us  assume that $\Z$ is   implemented  by the L\"uders instrument   $ \jj^L_x(\cdot) := \Z_x (\cdot) \Z_x $, which is repeatable due to sharpness of $\Z$. We accordingly  define the operations $\Phi_x^L (T) :=  \idchsys\otimes \jj_x^L(U T U^\dagger) $.  Note the following: $ \sum_x \onesys \otimes \Z_x = \sum_x \mathcal{Z}_x = \one$;   sharpness of $\Z$ and the Yanase condition implies ${\Phi_x^L}^*(H) = U^\dagger H (\onesys \otimes \Z_x) U \equiv U^\dagger (\onesys \otimes \Z_x) H U $;   and $\tr[AB] = \tr[BA]$. Therefore,  denoting the conditional change in energy in such a case as $\Delta \tilde \e^L(x)$,   taking the average  with respect to $p^\E_\rho (x)$ obtains 
\begin{align}\label{eq:conditional-first-law}
\avg{\Delta \tilde \e^L} &:= \sum_{x\in \xx} p^\E_\rho (x)   \Delta \tilde \e^L(x), \nonumber \\
& = \sum_{x\in \xx} \left( \tr[H {\Phi_x^L} (\rho \otimes  \xi) ]  - \frac{1}{2}\tr[\mathcal{Z}_x (H\rho \otimes  \xi  + \rho \otimes  \xi H )] \right), \nonumber \\
& = \left(\sum_{x\in \xx} \tr[{\Phi_x^L}^*(H) \rho \otimes  \xi ] \right) - \tr[H\rho \otimes  \xi ], \nonumber \\
& = \left( \sum_{x\in \xx} \tr[U^\dagger H (\onesys \otimes \Z_x) U \rho \otimes  \xi ] \right) - \tr[H\rho \otimes  \xi ], \nonumber \\
&= \tr[(U^\dagger H U - H) \rho \otimes  \xi ] = \ww.
\end{align}
Therefore,  we see that \eq{eq:conditional-first-law} is equivalent to the average first law shown in \eq{eq:average-first-law-1}; recall that when $\Z$ is sharp, satisfies the Yanase condition, and is implemented by a L\"uders instrument,  $\avg{\qq} = 0$. Indeed, in Ref. \cite{Mohammady2019c} this first law equality was used to identify the \emph{conditional work} with $\Delta \tilde \e^L(x)$:
\begin{align}\label{eq:conditional-work}
\tilde\ww(x)&:= \frac{\tr[H\Phi^L_x(\rho \otimes  \xi) ]}{p^\E_\rho (x)  }  - \frac{\frac{1}{2}\tr[\Phi^L_x (H\rho \otimes  \xi  + \rho \otimes  \xi H )]}{p^\E_\rho (x)  } \equiv \frac{\tr[H\Phi^L_x(\rho \otimes  \xi) ]}{p^\E_\rho (x)  }  - \frac{\frac{1}{2}\tr[\mathcal{Z}_x (H\rho \otimes  \xi  + \rho \otimes  \xi H )]}{p^\E_\rho (x)  }. 
\end{align}
If $[U, H]=\zero$, then we have ${\Phi^L_x}^*(H) =   H \mathcal{Z}_x = \mathcal{Z}_x H$, and so $\tilde\ww(x) =0$ for all $\rho \in \s(\hs)$ and $ x \in \xx$; not only will the  average conditional work $\avg{\tilde \ww}$ (or the unmeasured work $\ww$) vanish, but so too will the conditional work.   

While the heat was not considered in Ref. \cite{Mohammady2019c}, we may examine it here. Let us assume that $\Z$  is implemented by a general (not necessarily L\"uders) repeatable instrument $\jj$. Using \eq{eq:conditional-energy-change-tot} and \eq{eq:conditional-work},  we may  define the \emph{conditional} heat as $\tilde \qq(x) := \Delta \tilde \e(x) - \tilde \ww(x)$. However,   it becomes clear that $\tilde \qq(x)$  will no longer offer the same interpretation as objectification heat; the conditional heat reads  
\begin{align}
\tilde \qq(x)& =\frac{\tr[H \Phi_x(\rho \otimes  \xi )]}{p^\E_\rho (x)  } -   \frac{\tr[H \Phi^L_x(\rho \otimes  \xi )]}{p^\E_\rho (x)  } = \tr[H(\sigma_x - \sigma_x')] = \tr[H\sub{\aa} (\xi_x - \xi_x')],
\end{align}
where $\sigma_x:= \Phi_x(\rho \otimes  \xi )/p^\E_\rho (x)$ and $\sigma_x' := \Phi_x^L(\rho \otimes  \xi )/p^\E_\rho (x)$,   while   $\xi_x := \tr\sub{\s}[\sigma_x]$ and $\xi_x' := \tr\sub{\s}[\sigma_x'] $. The final equality  follows from the fact that the reduced state of the system is independent of how $\Z$ is implemented (see \eq{eq:instrument-dilation-2}). We see that contrary to the objectification heat, which is the change in energy due to objectification $ U(\rho \otimes  \xi ) U^\dagger \mapsto \sigma_x$, the conditional heat is now a counterfactual quantity, that is, it is given as the difference  between the expected energy of the actual objectified state $\xi_x$, and the expected energy of the counterfactual objectified state $\xi_x'$, obtained if $\Z$ \emph{were}  implemented by a L\"uders instrument; implementing $\Z$ by a L\"uders instrument implies a vanishing conditional heat.

\end{document}